\documentclass[11pt]{article}

\usepackage[nohyperref]{acl}

\usepackage{times}
\usepackage{latexsym}
\usepackage[T1]{fontenc}
\usepackage[utf8]{inputenc}
\usepackage{microtype}

\usepackage{graphicx}
\graphicspath{{figures/}}
\usepackage{amsmath,amssymb}
\usepackage{booktabs}
\usepackage{algorithm}
\usepackage{algorithmic}
\usepackage{tikz}
\usetikzlibrary{shapes.geometric, arrows, positioning}
\usepackage{float}
\usepackage{subcaption}
\usepackage{xcolor}
\PassOptionsToPackage{hyphens}{url}
\usepackage{url}
\newcounter{promptctr}[subsubsection]
\renewcommand{\thepromptctr}{\thesubsubsection.\arabic{promptctr}}

\newsavebox{\promptcontentbox}
\newcommand{\promptheading}[2]{%
  \refstepcounter{promptctr}%
  \par\medskip\noindent
  {\small\sffamily\bfseries Prompt~\thepromptctr: #1}\par\noindent{#2}\par\smallskip
}
\newenvironment{promptcontent}{%
  \noindent
  \begin{lrbox}{\promptcontentbox}%
  \begin{minipage}{\dimexpr\columnwidth-14pt}%
  \small
}{%
  \end{minipage}%
  \end{lrbox}%
  \fcolorbox{black!30}{black!4}{\usebox{\promptcontentbox}}%
  \par\medskip
}

\title{WARP: Wasserstein-Aligned RAG for Population Opinions}

\author{%
  Aman Singh Thakur \quad Aditya Agrawal \quad Alwarappan Nakkiran \quad Alex Karlsson \\[4pt]
  Amazon.com \\[4pt]
  \texttt{\{amanzing, aadag, nakkiran, nkarlss\}@amazon.com}
}

\begin{document}
\maketitle

\begin{abstract}
RAG systems are increasingly used to summarize what large collections of documents say. A user asks ``What do people think about X?'' and receives an answer that reads as consensus. But standard top-$k$ retrieval ranks documents by query similarity, not by how faithfully they represent the population, so minority views quietly disappear. Existing fixes fall short. Diversity re-rankers like MMR and DPP spread retrieved documents apart, but with no target distribution to aim for. Calibration methods based on KL or JS divergence do target one, yet treat opinion bins as unordered: confusing strong positive with strong negative costs no more than an adjacent-bin miss.

We introduce WARP, a family of post-retrieval algorithms that calibrate retrieved evidence to the population's opinion distribution. WARP first recovers underrepresented opinions that cosine ranking may bury, then uses Wasserstein-1 distance to select documents whose sentiment-intensity distribution matches the population target, capturing the ordinal structure ignored by KL and JS divergence. We develop three variants for dense, sparse, and variable candidate pools, trading off calibration quality and speed. Across three review domains spanning 35K documents, 156 queries, and 26 entities, WARP's domain-matched variants reduce distributional error by at least 43\% with sub-second latency. These gains carry through to generation: a five-judge LLM panel prefers WARP-generated answers in up to 86\% of decided comparisons at $k{=}5$.
\end{abstract}

\begin{figure*}[t!]
\centering
\includegraphics[width=\textwidth]{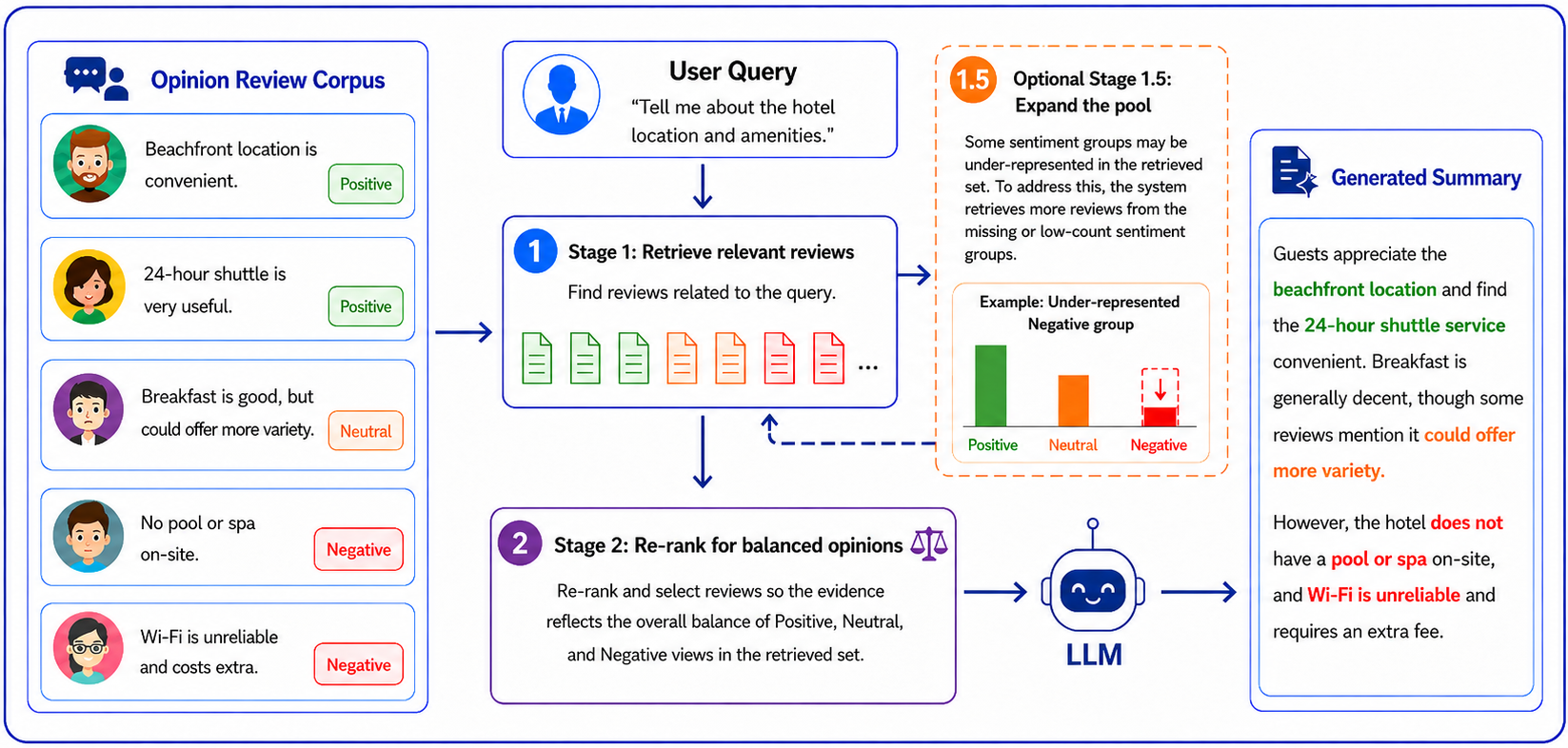}
\caption{WARP pipeline. \textbf{Stage~1}: semantic retrieval returns a Top-$N$ candidate pool. \textbf{Stage~1.5}: deficit-aware pool expansion recovers under-represented sentiment poles via re-retrieval (\emph{Entity-Gated} or \emph{Adaptive Expansion}; \S\ref{sec:pool-expansion}; self-bypasses when the pool is already balanced). \textbf{Stage~2}: $W_1$ re-ranking selects the final $k$ documents whose empirical opinion distribution matches the population target $P_{\text{pop}}$ (\emph{$W_1$ Minimizer} for dense pools, \emph{$W_1$-MMR} for sparse pools, or \emph{WassRank OT} as a tuning-free fallback; \S\ref{sec:reranking}). Pool density drives the variant choice (decision matrix in Table~\ref{tab:deployment}).}
\label{fig:pipeline}
\end{figure*}

\section{Introduction}

Users used to browse reviews themselves - clicking through results, weighing contradictions, building a mental picture. Noisy, but transparent: you could see disagreement. RAG systems~\citep{lewis2020rag} replace this with a single synthesized answer, but a new failure mode emerges. Standard top-$k$ retrieval selects evidence by query similarity, without factoring spread of opinions in the corpus. Take a hotel where 60\% of reviewers are satisfied and 40\% complain about noise. Retrieval over-represents the majority because similarity scoring favors the dominant cluster. Every claim in the resulting summary traces to a real review, yet the user never learns that two in five guests had a bad experience. This leads to distributional distortion. A synthesized response from a skewed sample would read factual but would not present the whole picture.

For opinion queries, uncertainty is aleatoric: it reflects genuine disagreement. A faithful system must preserve the distribution without collapsing it. \citet{agrawal2026opinionrag} formalize this distinction; \citet{nayeem2025opiniorag} build a scalable opinion summarization pipeline but do not optimize the retriever for distributional fidelity. \textbf{What's missing is the engineering: in a production stack with sub-second latency budgets, how do you select a small evidence set whose opinion distribution actually matches the population?}

A standard cosine retriever maximizes query similarity, producing relevant but homogeneous evidence - four glowing reviews when two would suffice. Diversity re-rankers (MMR~\citep{carbonell1998mmr}, DPP~\citep{kulesza2012dpp}) push back by maximizing pairwise spread, which helps until each sentiment bin has one representative; past that point they quietly revert to relevance ordering. KL and JS calibration~\citep{steck2018calibrated,dang2012pm2} come closer by explicitly optimizing against a target distribution. But they remain categorical. Confusing ``positive'' with ``negative'' costs the same as confusing ``positive'' and ``neutral''. Opinions are ordinal and we need a metric which supports this.

Wasserstein-1 respects this ordering. Its closed-form CDF computation ($O(m)$ for $m$ sentiment bins) makes per-candidate evaluation cheap enough for runtime re-ranking without offline precomputation. We built WARP (Figure~\ref{fig:pipeline}) around this in two stages. First, a deficit-aware pool expansion pass compares the retrieved distribution against the population target, identifies under-represented poles, and issues targeted re-retrievals only where gaps exist. Then a greedy re-ranker selects the final evidence set by minimizing $W_1$ distance to $P_{\text{pop}}$. The pipeline instantiates in three variants (Section~\ref{sec:reranking}): the Minimizer greedily selects from entity-matched candidates - effective for dense pools; for sparser conditions, $W_1$-MMR blends relevance with calibration, and WassRank OT solves the full slot-assignment as rectangular optimal transport requiring no per-domain tuning.

Our contributions:
\begin{enumerate}
    \item \textbf{Pool expansion that recovers what cosine retrieval buries.} Entity-gated and adaptive re-retrieval recover under-represented opinions with near-complete entity coverage. Self-bypasses at zero cost when already balanced.
    \item \textbf{$W_1$ re-rankers substantially reduce distributional error.} A greedy minimizer that matches opinion distribution for entity dense domains and the hybrid variants trade opinion diversity for relevance on sparse domains.
    \item \textbf{Deployment characterization across 3 domains.} Pool density determines algorithm choice. All variants add under 310\,ms re-ranking latency (p99) and generalize across labeling methods and index types.
\end{enumerate}

\begin{figure*}[t!]
\centering
\includegraphics[width=\textwidth]{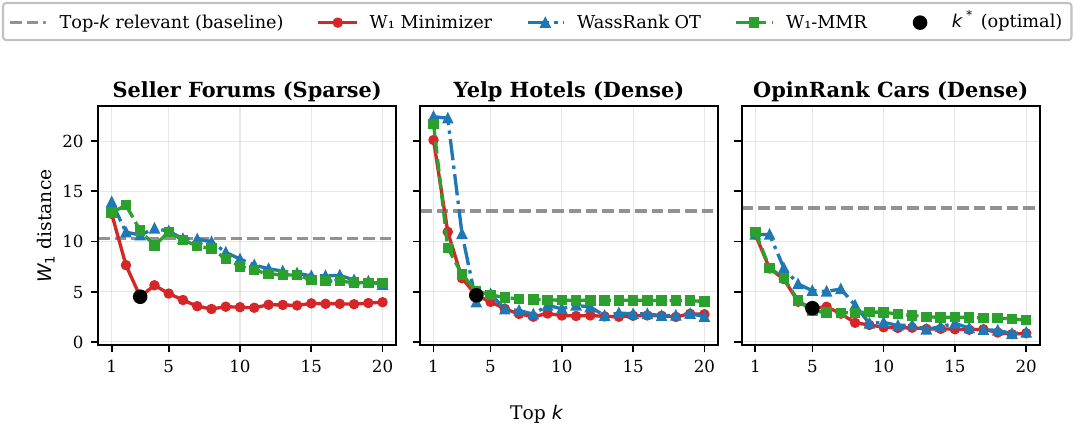}
\caption{$W_1$ re-rankers (no re-retrieval): domain-matched variants reduce distributional error at least 43\% vs.\ Top-$k$ (gray dashed) --- 43\% Seller Forums ($W_1$-MMR), 80\% Yelp ($W_1$ Minimizer), 69\% OpinRank ($W_1$ Minimizer); see Table~\ref{tab:main}. Curves converge at $k^*{=}3$--$5$ documents (black dot; Appendix~\ref{app:adaptive-k}) as most entities' opinion mass concentrates in 2--3 of 7 bins. $N{=}200$.}
\label{fig:w1_vs_k}
\end{figure*}

\section{Related Work}

\paragraph{Retrieval diversity.}
MMR \citep{carbonell1998mmr} penalizes redundancy via pairwise dissimilarity; DPP \citep{kulesza2012dpp} maximizes determinantal spread; xQuAD \citep{santos2010xquad} diversifies over query intents. More recent variants-BQP \citep{lu2026bqp}, AdaGReS \citep{peng2025adagres}, MUSS \citep{nguyen2025muss}-refine these ideas in various ways (see \citet{wu2024diversity_survey} for a survey). However, diversity is the wrong objective here as the signal runs out, once every sentiment bin has one representative (5--7 selections).

\paragraph{Optimal transport (OT) in IR.}
Wasserstein distances~\citep{peyre2019computational} have appeared in IR before-Word Mover's Distance \citep{kusner2015wmd} for document similarity, WassRank \citep{yu2019wassrank} as a listwise training loss, OTExtSum \citep{tang2022otextsum} for extractive summarization, Wasserstein coresets \citep{claici2018coresets} for data summarization. However, all these function at training time, over generic content. To our knowledge, none use Wasserstein as a \emph{runtime} selection objective conditioned on an opinion distribution.

\paragraph{Opinion-aware retrieval.}\looseness=1
\citet{agrawal2026opinionrag} audit 30+~RAG benchmarks and find a striking gap: only one addresses opinion synthesis. They formalize the epistemic/aleatoric distinction - the posterior converges to the population distribution $P_{\text{pop}}(\theta)$ for opinion queries - and propose a three-term objective combining coverage ($W_2$), fidelity, and demographic fairness. However, their work is limited to 2 domains without practical retrieval strategies. \citet{nayeem2025opiniorag} solve a different problem entirely: generating opinion highlights from thousands of reviews via retrieve-then-synthesize with AOS-triplet verification. Their retriever is standard semantic search without targeting distributional fidelity. We instantiate the \citet{agrawal2026opinionrag} coverage objective using $W_1$ instead of $W_2$ - its closed-form CDF difference enables $O(m)$ per-candidate evaluation at runtime - and test across three domains.

\paragraph{Calibrated selection and diversity methods.}
Calibrated recommendation via KL minimization \citep{steck2018calibrated} and proportional slot allocation (PM-2; \citealt{dang2012pm2}) do address distributional fidelity across facets - but assume categorical labels. Fairness-aware ranking \citep{singh2018fairness,zehlike2017fair,oesterling2024proportional} enforces demographic constraints without modeling ordinal opinion at all. On the sentiment side, \citet{aktolga2013sentiment} applies bias modes to retrieval, aspect-level summarization \citep{angelidis2018summarizing,jiang2023subsumm} diversifies across topical facets. The methods we surveyed all use categorical penalty structures that do not exploit ordinal distance between opinion bins. WARP addresses this gap: an ordinal $W_1$ ground cost applied at query time to opinion-distribution matching in RAG, drawing on the calibrated-selection tradition~\citep{steck2018calibrated} and OT-in-ranking~\citep{yu2019wassrank} while adapting both to a training-free runtime re-ranker across three review domains.

\section{Methodology}

Below we describe the experimental setup (Sections~\ref{sec:datasets}--\ref{sec:baselines}), pool expansion (Section~\ref{sec:pool-expansion}), and $W_1$ re-ranking (Section~\ref{sec:reranking}) for stress-testing WARP.

\subsection{Datasets}
\label{sec:datasets}

3 review corpora give us the spread we need across different domains:

\begin{itemize}
\item \textbf{Amazon Seller Forums}\footnote{\url{https://sellercentral.amazon.com/seller-forums}} (${\sim}$8K documents). Retrieved via Amazon Bedrock Knowledge Bases~\citep{aws2024bedrockkb} with hybrid search.
\item \textbf{Yelp Open Dataset}~\citep{yelp2024dataset} (${\sim}$14K hotel reviews). Retrieved via local FAISS~\citep{johnson2019faiss} + MiniLM-L6~\citep{wang2020minilm,reimers2019sentencebert}.
\item \textbf{OpinRank}~\citep{ganesan2012opinrank} (${\sim}$13K automotive reviews). Retrieved via local FAISS + MiniLM-L6.
\end{itemize}

Entity selection follows \citet{agrawal2026opinionrag}: LLM-based entity extraction, diversity filtering (Shannon entropy $H \geq 0.6$, minority sentiment ${\geq}10\%$), then ranking by entropy. 

We use six questions per entity (2 breadth, 2 polar, 2 segment; templates in Appendix~\ref{app:queries}) which gives us 156 total queries, across all domains. Entity and sentiment-intensity labels come from a single LLM pass per document (prompt in Appendix~\ref{app:prompts:extraction}), mapped to a 7-bin ordinal scale $\mathcal{S} = \{-30, -20, -10, 0, +10, +20, +30\}$ (Table~\ref{tab:si-scale}, Appendix~\ref{app:datasets}). 

Full dataset statistics appear in Table~\ref{tab:datasets} (Appendix~\ref{app:datasets}).

Defaults: $N{=}200$ candidate pool, $k{=}20$ output.

\subsection{Baselines}
\label{sec:baselines}

Five baselines. Top-$k$ is plain cosine-similarity retrieval. MMR~\citep{carbonell1998mmr} penalizes semantic redundancy ($\lambda{=}0.5$). DPP~\citep{kulesza2012dpp} maximizes determinantal spread over 1D sentiment-intensity features. OpinionMMR penalizes opinion-bin distance rather than embedding distance. KL (JS) Minimizer implements \citet{steck2018calibrated}'s calibrated-recommendation objective with Jensen--Shannon divergence - identical to our $W_1$ Minimizer but with an order-agnostic metric (Appendix~\ref{app:kl-ablation}).

\paragraph{Metrics.} Distributional fidelity: Wasserstein-1 distance $W_1$ (lower is better). Entity relevance: Entity Match rate (EM\%, hereafter EM; fraction of selected documents matching the queried entity, higher is better). We also report re-ranking latency in milliseconds. For generation evaluation, Cohen's $\kappa$~\citep{cohen1960kappa} measures inter-judge alignment.

\subsection{Pool Expansion via Re-Retrieval}
\label{sec:pool-expansion}

For entity sparse domains, the relevance-based retrieval step typically returns a substantial fraction of documents that do not discuss the queried entity - due to near-misses leaked in from other entities sharing the same index. We tackle this with two expansion strategies (full algorithms in Appendix~\ref{app:pool-algorithms}):

\paragraph{Entity-Gated Re-retrieval.} Pass~1 pulls the standard top-$N$ by relevance. Pass~2 then digs into the tail of the ranked list, fishing out documents that match the queried entity but fell below the cutoff, ranking them by opinion extremity (absolute sentiment-intensity score - strongly negative reviews surface before mild ones). The passes are merged and deduplicated.

\paragraph{Adaptive Expansion.} We compare the pool's opinion breakdown against $P_{\text{pop}}$. If any bin is under-represented (pool fraction below threshold of the population fraction), a targeted retrieval fires for entity-matched documents in the deficit bin using pole-biased queries (Prompt~\ref{prompt:pole}). If the pool already mirrors the target, nothing happens - zero overhead (Algorithm~\ref{alg:adaptive-expansion}).

\begin{table*}[t]
\centering
\small
\caption{Main results: Diversity baselines and calibrated methods vs WARP re-rankers. $W_1{\downarrow}$: distributional distance to the population opinion target. EM\%${\uparrow}$: fraction of selected documents matching the queried entity. EG/AE = pool expansion strategies (Entity-Gated / Adaptive Expansion). $N{=}200$ candidates, $k{=}20$ output. Paired Wilcoxon vs.\ Top-$k$: $^{***}p{<}0.001$, $^{**}p{<}0.01$, $^{*}p{<}0.05$. Bold: best per column. $^\dagger$Shared FAISS index across all OpinRank entities (EM=82.9\%); entity-specific indices give $W_1{=}0.95$ with 100\% EM (Appendix~\ref{app:lambda}).}
\label{tab:main}
\begin{tabular}{l cc cc cc c}
\toprule
& \multicolumn{2}{c}{\textbf{Seller Forums}} & \multicolumn{2}{c}{\textbf{Yelp Hotels}} & \multicolumn{2}{c}{\textbf{OpinRank Cars}} & \textbf{Latency} \\
\cmidrule(lr){2-3} \cmidrule(lr){4-5} \cmidrule(lr){6-7}
\textbf{Method} & $W_1{\downarrow}$ & EM\% & $W_1{\downarrow}$ & EM\% & $W_1{\downarrow}$ & EM\% & \textbf{(ms)} \\
\midrule
Top-$k$ (baseline) & 10.33 & 42.9 & 13.60 & 40.1 & 13.33 & 27.0 & \textbf{0} \\
MMR\textsuperscript{*} & 10.31 & 43.5 & 8.67 & 14.2 & 9.64 & 11.3 & 12 \\
DPP\textsuperscript{***} & 11.69 & 41.2 & 5.63 & 78.8 & 7.38 & 69.2 & 45 \\
OpinionMMR\textsuperscript{**} & 10.44 & 43.5 & 8.59 & 12.4 & 10.02 & 10.6 & 12 \\
KL (JS) Minimizer\textsuperscript{*} & 8.87 & 51.4 & 3.00 & 93.8 & 4.54 & 85.8 & \textbf{9} \\
\midrule
$W_1$-MMR\textsuperscript{***} & 5.86 & 44.6 & 4.02 & 19.4 & \textbf{3.41} & 21.8 & 154 \\
$W_1$ Minimizer\textsuperscript{***} & 8.84 & 51.4 & 2.76 & 93.8 & 4.14 & \textbf{85.8} & \textbf{9} \\
WassRank OT\textsuperscript{***} & \textbf{5.71} & 44.4 & 2.52 & 92.2 & 3.87$^\dagger$ & 82.9 & 89 \\
EG + $W_1$ Min\textsuperscript{***} & 8.59 & \textbf{63.5} & \textbf{1.52} & \textbf{99.2} & 4.14 & 85.8 & 13 \\
AE + $W_1$ Min\textsuperscript{***} & 8.59 & \textbf{63.5} & 1.66 & 95.9 & 4.14 & 85.8 & \textbf{8} \\
\bottomrule
\end{tabular}
\end{table*}

\subsection{Wasserstein Re-Ranking Algorithms}
\label{sec:reranking}

\begin{algorithm}[t!]
\caption{$W_1$ Minimizer}
\label{alg:w1min}
\begin{algorithmic}[1]
\REQUIRE\hspace{-0.5em}Pool $C$,entity $e$,target $P_{\text{pop}}$,output size $k$
\STATE $C_e \leftarrow \{d \in C : \text{entity}(d) = e\}$
\STATE $S \leftarrow \emptyset$
\WHILE{$|S| < k$}
    \STATE $d^* \leftarrow \arg\min_{d \in C_e \setminus S} W_1(\text{EmpDist}(S \cup \{d\}), P_{\text{pop}})$
    \STATE \hspace{1em} ties broken by relevance score $\downarrow$
    \STATE $S \leftarrow S \cup \{d^*\}$
\ENDWHILE
\RETURN $S$
\end{algorithmic}
\end{algorithm}

\paragraph{Problem Formulation.}
For opinion queries, a faithful answer must reflect how viewpoints distribute along an ordinal axis - one where positions have a natural ordering and distances between them reflect severity of disagreement. The re-ranking objective follows directly: select $k$ documents whose empirical distribution along this axis minimizes transport distance to the target distribution $P_{\text{pop}}$. We instantiate the axis as sentiment-intensity (SI), a 7-bin ordinal scale $\mathcal{S} = \{-30, -20, \ldots, +30\}$. At indexing time each document $d$ receives an SI label; per entity $e$ we derive $P_{\text{pop}}(s)$ as the corpus-observed fraction of documents for $e$ in each bin, or as an externally-specified target (star ratings, survey-calibrated priors) when one is available --- $P_{\text{pop}}$ is an input to the re-ranker, not a claim about the true underlying population (§Limitations; tolerance to partial estimates in Section~\ref{sec:deployment}). Given a candidate pool $C$ of $N$ documents, pick $k \ll N$ into result set $S$ minimizing:
\begin{equation}
W_1(P, Q) = \sum_{i=1}^{m-1} \left| \text{CDF}_P(s_i) - \text{CDF}_Q(s_i) \right| \cdot \Delta s
\label{eq:w1}
\end{equation}
where CDF is the cumulative distribution function, $m {=} 7$ bins, and $\Delta s {=} 10$ is the bin spacing. The ordinal structure matters: confusing $+30$ with $-30$ costs $W_1^{\max} {=} 60$, while adjacent-bin errors cost only 10. Evaluation runs in $O(m)$ per candidate; we observe empirically that rank ordering is preserved under $W_2$ (Appendix~\ref{app:w2}).

We implemented three variants that minimize this objective. They differ in how they cope with entity pool density and the relevance-calibration tradeoff:

\paragraph{$W_1$ Minimizer.}
Filter the pool to entity-matched candidates, then greedily add whichever document pulls the empirical distribution closest to $P_{\text{pop}}$, breaking ties by relevance score (Algorithm~\ref{alg:w1min}). Needs a dense pool where most candidates match the queried entity.

\paragraph{$W_1$-MMR.}
Built for sparse pools where entity-matched candidates alone cannot fill $k$ slots. This variant works on the full candidate set - no entity filtering - and scores each candidate by blending retrieval relevance with the calibration gain it would contribute:
\begin{equation}
\text{score}(d) = \lambda \cdot \text{rel}(d) + (1-\lambda) \cdot \frac{\Delta W_1}{W_1^{\text{cur}}}
\label{eq:w1mmr}
\end{equation}
$\text{rel}(d)$ is the original retrieval score, $\lambda$ controls the relevance-calibration blend, and dividing by $W_1^{\text{cur}}$ normalizes to percentage improvement so the calibration term does not vanish as the distribution tightens. Full algorithm in Appendix~\ref{app:w1mmr}.

\paragraph{WassRank OT.}
Instead of greedy selection, WassRank~\citep{yu2019wassrank} solves the assignment globally: allocate $k$ slots proportionally to $P_{\text{pop}}$, then find the minimum-cost candidate-to-slot bijection where cost blends normalized ordinal distance with a relevance penalty. This minimizes blended transport cost to a proportional discretization of $P_{\text{pop}}$, which empirically tracks $W_1$ (Appendix~\ref{app:lambda}). We solve the rectangular assignment~\citep{crouse2016rectangular} in $O(k^2 \cdot N)$. Originally a listwise training loss; we repurpose it here as an inference-time re-ranker with no learned parameters and call it WassRank OT.

\section{Results}

Four questions drove our experiments: does it work, does the metric choice matter, do retrieval gains actually reach the user, and which variant belongs where? Sections~\ref{sec:main-results}--\ref{sec:deployment} take them in order.

\subsection{Main Results}
\label{sec:main-results}

In Table~\ref{tab:main}, the gains are large and consistent, concentrating on entities with the most skewed baseline distributions (per-entity breakdowns in Appendix~\ref{app:per-entity}). Our domain-matched $W_1$ re-ranking algorithms (Section~\ref{sec:reranking}) cut distributional error (Equation~\ref{eq:w1}) by at least 43\% relative to Top-$k$ --- 43\% on sparse Seller Forums ($W_1$-MMR), 80\% on dense Yelp ($W_1$ Minimizer, rising to 89\% with pool expansion), and 69\% on OpinRank ($W_1$ Minimizer). Entity relevance tracks along: pairing Entity-Gated re-retrieval (Section~\ref{sec:pool-expansion}) with $W_1$ Minimizer pushes Yelp EM\% from 40.1\% (Top-$k$ baseline) to 99.2\% (post-reranking). As a sanity check, 5\% Gaussian perturbation of retrieval scores yields only 2--5\% $W_1$ improvement, non-significant. This confirms that the score structure drives these results and not the arbitrary reshuffling. (Appendix~\ref{app:noise-control})

\subsection{The Ordinal Metric Makes a Measurable Difference}
\label{sec:ordinal}

The metric matters. Swapping JS divergence for $W_1$ while keeping the same greedy loop and entity filtering cuts error by 8.7\% on Yelp and 9.7\% on OpinRank (Appendix~\ref{app:kl-ablation}). JS penalizes a $+30 \leftrightarrow -30$ confusion no more than an adjacent-bin slip; $W_1$ charges six times more. Entity filtering alone accounts for 45.7\% $W_1$ reduction on Yelp; $W_1$ Minimizer adds 62.6\% beyond that, and on entity sparse domains, like Seller Forums, $W_1$-MMR adds 37.7\% beyond the entity-filter baseline (Appendix~\ref{app:entity-control}). We also stress-tested under noise: corrupting both SI labels and $P_{\text{pop}}$ simultaneously, $W_1$ Minimizer never falls below Top-$k$ even at 50\% label error on dense pools, and still outperforms Top-$k$ by 72\% at 20\% Dirichlet perturbation (Appendix~\ref{app:robustness}).

\subsection{Retrieval Gains Propagate to Generation}
\label{sec:generation}

Do retrieval gains actually reach users? We generated answers from each method's retrieved evidence (prompt in Appendix~\ref{app:prompts:generation}) and ran blind pairwise comparisons against Top-$k$ using a 5-judge LLM panel (Claude Sonnet 4~\citep{anthropic2025claude}, Llama 3.3 70B~\citep{meta2024llama33}, Mistral Large 3~\citep{mistral2025large3}, Amazon Nova Pro~\citep{agi2025amazonnova}, DeepSeek v3.2~\citep{deepseek2025v32}) with position control~\citep{zheng2024judging} and 3/5 majority vote (Appendix~\ref{app:judge-validation}). All five judges from independent model families agree directionally at $\kappa_d{=}0.61$--$1.00$; the random baseline correctly shows no significance.

At $k{=}5$, calibration methods win 70--89\% of decided comparisons (Table~\ref{tab:gen-cross}). At $k{=}10$ this drops to 73--88\% (Table~\ref{tab:gen-cross-k10}) as larger context lets the LLM average out distributional imbalance (Appendix~\ref{app:generation}). A prompt ablation (Appendix~\ref{app:prompt-ablation}) disentangles two evaluation criteria: when judges assess \emph{proportional accuracy} (does the answer reflect how common each view is?), $W_1$ Minimizer leads DPP by 8\,pp and matches an oracle stratified-sampler within 1\,pp (Table~\ref{tab:prompt-ablation}); under pure \emph{viewpoint coverage} (does it mention all perspectives?), the gap narrows to parity. $W_1$'s proportional advantage persists through generation, not just retrieval.

\paragraph{Metric isolation.}
On sparse pools, $W_1$-MMR appears to lose to KL(JS)-MMR (Appendix~\ref{app:kl-ablation}), but the comparison is confounded: KL(JS)-MMR selects 1.7--2$\times$ more entity-matched documents than $W_1$-MMR on Yelp and OpinRank. Once we control for EM\%, the Minimizer comparison flips: $W_1$ wins on both domains (Appendix~\ref{app:em-confound}), and a single-judge generation evaluation confirms the same direction, with $W_1$ Minimizer leading KL(JS) Minimizer by 7\,pp Fair\% cross-domain (Appendix~\ref{app:metric-isolation}).

Two other patterns are worth noting. First, semantic-diversity baselines fade as $k$ grows: MMR and OpinionMMR Fair\% collapse to 29--33\% at $k{=}20$, while distributional methods keep gaining (WassRank OT: 70\%$\to$73\%$\to$92\% across $k{\in}\{5,10,20\}$; Appendix~\ref{app:gen-k20}). Second, DPP and $W_1$ optimize different targets --- bin diversity vs.\ bin ratios --- which happen to align at $k{=}5$ with 2--3 active bins but diverge at $k{=}20$. On OpinRank, near-uniform $P_{\text{pop}}$ compresses ordinal distances, so the KL/$W_1$ gap narrows; the ordinal advantage is largest where distributions are skewed (Yelp, Seller Forums).

\begin{table}[t]
\centering
\small
\caption{Generation evaluation ($k{=}5$): 5-judge majority vote, position-controlled blind pairwise vs.\ Top-$k$ ($N{=}156$ queries). Fair\% = W/(W+L). Win\% = W/$N$. All approaches significant at $p{<}0.001$ (sign test) except Random. $k{=}10$ in Appendix~\ref{app:generation}.}
\label{tab:gen-cross}
\resizebox{\columnwidth}{!}{%
\begin{tabular}{lccc}
\toprule
\textbf{Approach} & \textbf{W/L/T} & \textbf{Fair\%} & \textbf{Win\%} \\
\midrule
Random & 16/36/104 & 31\% & 10\% \\
MMR$^{*}$ & 39/8/109 & 83\% & 25\% \\
DPP$^{*}$ & 54/7/95 & \textbf{89\%} & 35\% \\
\midrule
WassRank OT$^{*}$ & 66/28/62 & 70\% & \textbf{42\%} \\
W$_1$-MMR$^{*}$ & 44/10/102 & 81\% & 28\% \\
W$_1$ Minimizer$^{*}$ & 51/8/97 & \textbf{86\%} & 33\% \\
\bottomrule
\end{tabular}%
}
\end{table}

\subsection{Deployment Characterization}
\label{sec:deployment}
\label{sec:pool-density}

Which algorithm to pick comes down to one number: entity match percentage (Table~\ref{tab:deployment}). Dense pools, where all retrieved documents discuss the query entity, are easy. When EM${\geq}85\%$, $W_1$ Minimizer delivers 69--81\% reduction, statistically indistinguishable from oracle stratified sampling. Entity-Gated re-retrieval pushes Yelp further ($W_1$: 2.76$\to$1.52, EM\%: 93.8$\to$99.2); on OpinRank both expansion strategies detect no deficit and self-bypass. Sparse entity pools are a different story. On Seller Forums (Top-$k$ baseline EM${\sim}$40\%; $W_1$ Minimizer reaches ${\sim}$51\% post-reranking), $W_1$-MMR scores all 200 candidates regardless of entity match, reaching 43\% reduction while maintaining 81\% generation propagation (Table~\ref{tab:gen-cross}). This outperforms WassRank OT (70\%) despite comparable retrieval-side $W_1$, making $W_1$-MMR the preferred hybrid for sparse and variable pools. Re-ranking latency stays under 310\,ms across the board; end-to-end depends on retrieval infrastructure (Appendix~\ref{app:latency}). For dense pools $N{=}100$ suffices; $W_1$-MMR benefits from larger $N$ on sparse pools (Appendix~\ref{app:n-sensitivity}).

\begin{table}[t]
\centering
\small
\caption{Deployment decision matrix. Pool density determines algorithm choice; all methods SLA-compliant ($N{=}200$, $k{=}20$, $m{=}7$ bins). $|C_e|$ = entity-matched candidates. Noise tolerance details in Appendix~\ref{app:robustness}.}
\label{tab:deployment}
\resizebox{\columnwidth}{!}{%
\begin{tabular}{@{}llcl@{}}
\toprule
\textbf{Pool Condition} & \textbf{Method} & \textbf{ms} & \textbf{Complexity} \\
\midrule
Dense (EM${\geq}$85\%) & $W_1$ Minimizer & 9 & $O(|C_e| \cdot k \cdot m)$ \\
Sparse (baseline EM${\sim}$40\%) & $W_1$-MMR & 154 & $O(N \cdot k \cdot m)$ \\
Variable / unknown & EG + $W_1$ Min & 13 & $O(|C_e| \cdot k \cdot m)$ \\
\bottomrule
\end{tabular}%
}
\end{table}

An exact $P_{\text{pop}}$ turns out to be unnecessary: fifty entity-matched documents preserve 74--98\% of oracle improvement; smoothing with a domain prior ($\tau{=}50$) matches the oracle's harm rate. Cold-start entities fall back to this prior while still beating Top-$k$ (Appendix~\ref{app:ppop-estimation}). The pipeline runs unchanged across managed cloud vector stores and local FAISS indices.

\subsection{Independent-Target Validation}
\label{sec:star-validation}

A natural concern is circularity: our LLM-derived sentiment-intensity (SI) labels define both the population target $P_{\text{pop}}$ used at inference and the $W_1$ metric applied to the retrieved evidence. If the labeler had systematic biases, calibrating and measuring against it could inflate the numbers. To rule this out, we cross-validate against Yelp's native 1--5 star ratings on all 13{,}533 reviews. Stars are ordinal, human-authored at review time, and participate in \emph{no part} of the WARP pipeline --- neither retrieval, nor re-ranking, nor the $P_{\text{pop}}$ target. Mapped to the same 7-bin SI scale, star-derived and LLM-derived labels correlate at Spearman $\rho{=}0.881$ ($p{<}10^{-300}$) and Pearson $r{=}0.908$, with 81.3\% same-valence agreement and MAE 7.2 SI units --- less than one bin (Appendix~\ref{app:star-validation}).

Correlation on individual labels is necessary but not sufficient; a labeler could still consistently mis-estimate the entity-level $P_{\text{pop}}$. We therefore recompute $P_{\text{pop}}$ from star ratings alone for each of the 10 Yelp entities and evaluate WARP's retrievals against this fully independent target. Mean $W_1$ gap between star-derived and LLM-derived $P_{\text{pop}}$: 2.88 (Appendix Table~\ref{tab:star-ppop}). Triangle inequality then gives $W_1(\text{WARP}, \text{star-}P_{\text{pop}}) \leq 5.64$ and $W_1(\text{Top-}k, \text{star-}P_{\text{pop}}) \geq 10.72$, i.e., a worst-case reduction of at least 47.4\% \emph{against a target our labels never touched}. This complements the mislabel sensitivity study (Appendix~\ref{app:sec:robustness}), where $W_1$ Minimizer never crosses the Top-$k$ baseline on Yelp even at 50\% label corruption, along with the FDR-corrected significance testing and entity-clustered bootstrap in Appendix~\ref{app:sec:stat-validity}. The gains are not an artifact of the labeler or of any single evaluation protocol.

\section{Conclusion}

As enterprise document corpora grow, RAG systems have become the default interface between users and organizational knowledge, returning relevant evidence in sub-second re-ranking latencies. Yet when the underlying corpus is diverse and highly opinionated, relevance-ranked retrieval selects only for topical similarity; the retrieved set may not reflect population opinions proportionally, producing summaries that appear grounded but misrepresent the balance of views.

WARP closes this gap by re-ranking retrieved evidence against the target population distribution, requiring no model fine-tuning, no retriever changes, and no additional inference calls. The result is at least a 43\% reduction in distributional distance to the population target ($W_1$), all within a 310\,ms latency envelope suitable for production deployment.

\section*{Limitations}

\paragraph{Offline evaluation under production constraints.}
All reported results come from offline experiments, engineered to satisfy the sub-310\,ms \emph{re-ranking} budget typical of a large-scale e-commerce RAG setting (end-to-end depends on retrieval infrastructure; Appendix~\ref{app:latency}). We therefore characterize retrieval and generation-side fidelity, not live user outcomes: whether proportionally faithful summaries measurably shift user trust, engagement, or decision quality under live traffic is untested, and offline $W_1$ reductions need not map one-to-one onto user-perceived faithfulness. A controlled online A/B test is the natural next step.

\paragraph{Corpus bias vs.\ retrieval bias.}
$P_{\text{pop}}$ is \emph{corpus-observed}, not a true population target: review corpora are self-selected, motivated writers with strong positive or negative experiences are over-represented, and some segments never write at all. WARP removes the \emph{retrieval-induced} bias --- Top-$k$ layers cosine-similarity skew on top of the underlying corpus bias --- but does not remove the corpus bias itself. Faithful matching over a biased sample can lend false authority: the summary reflects the writers' distribution, not the underlying population's. Because $P_{\text{pop}}$ is an \emph{input} to the re-ranker rather than something it learns, external signals (native star ratings, demographic priors, survey-calibrated distributions) can be plugged in directly to reweight the target --- a natural extension requiring no architectural change.

\paragraph{Pre-computed labels required.}
Per-document sentiment-intensity annotations are needed at indexing time. A single LLM pass suffices, and 50--100 labels per entity are enough - but the cost--quality tradeoff of labeling strategies is not evaluated.

\paragraph{Scale and Temporal Drift.}
156 queries across 26 entities; paired Wilcoxon provides adequate power, but scaling beyond 14K documents is untested. $P_\text{pop}$ degrades gracefully under perturbation (72\% better than Top-$k$ at $\varepsilon{=}0.2$) but temporal drift is not addressed.

\paragraph{Single ordinal axis; multi-issue opinions.}
WARP operates on a single ordinal axis (sentiment intensity). Multi-issue opinion spaces --- e.g., a review that praises price but criticizes support, or patient-experience, employee-engagement, and multi-issue polling settings where several stance dimensions matter jointly --- would require multi-marginal optimal transport, a natural extension we do not evaluate here.

\paragraph{Domain scope and generation.}
All domains are product/service reviews requiring entity-anchored opinions and an ordinal scale. Generation significance is driven by Yelp (97\% decided) and OpinRank (94\%); Seller Forums differentiates weakly (54\%) due to sparse pools. A prompt ablation (Appendix~\ref{app:prompt-ablation}) shows reported gaps are criterion-dependent.

\section*{Ethics Statement}

Distributional fidelity proportionally surfaces minority views, including potentially harmful ones. A content-safety filter should gate re-ranker output: excluded opinions are removed from both $P_{\text{pop}}$ and the candidate pool \emph{before} calibration. Because re-ranking is post-retrieval, it is compatible with any upstream safety filter without modification.

\section*{Use of AI Writing Assistance}

The research (design, experiments, analysis, and writing) is the authors' own work. We used Claude (Anthropic) as a proofreader and sounding board: it flagged unclear phrasing, suggested structural edits, and helped scaffold parts of the experimental code, all of which we reviewed and revised ourselves. Figure~1 was generated with ChatGPT (OpenAI) to give readers a quick visual overview of the pipeline; we verified its accuracy. No AI system produced research claims or final prose.

\bibliography{custom}

\appendix
\appendix

\raggedbottom
\renewcommand{\topfraction}{0.92}
\renewcommand{\bottomfraction}{0.92}
\renewcommand{\textfraction}{0.05}
\renewcommand{\floatpagefraction}{0.88}
\setlength{\textfloatsep}{6pt plus 2pt minus 2pt}
\setlength{\floatsep}{6pt plus 2pt minus 2pt}
\setlength{\intextsep}{6pt plus 2pt minus 2pt}
\setlength{\dbltextfloatsep}{6pt plus 2pt minus 2pt}
\setlength{\dblfloatsep}{6pt plus 2pt minus 2pt}
\setlength{\abovecaptionskip}{4pt}
\setlength{\belowcaptionskip}{2pt}

\section{Experimental Setup}
\label{app:sec:setup}

\subsection{Dataset Details}
\label{app:datasets}

\paragraph{Opinion Scale.} Sentiment-Intensity (SI) is mapped to a discrete ordinal scale, shown in Table~\ref{tab:si-scale}.

\begin{table}[H]
\centering
\small
\caption{Sentiment-Intensity (SI) mapping to the 7-bin ordinal scale. \emph{Mixed} labels (praise-and-criticism co-occurring in one review) collapse to the neutral bin; they account for 4--8\% of extractions across domains and $W_1$ Minimizer never crosses the Top-$k$ baseline on Yelp even at 50\% label corruption (§\ref{app:robustness}), so this pooling is bounded in impact. Multi-issue axes require multi-marginal transport --- outside our current scope (§Limitations).}
\label{tab:si-scale}
\begin{tabular}{lll}
\toprule
\textbf{Sentiment} & \textbf{Intensity} & \textbf{SI Score} \\
\midrule
Positive & High / Med / Low & +30 / +20 / +10 \\
Neutral & Any & 0 \\
Mixed    & ---              & 0 \\
Negative & High / Med / Low & $-30$ / $-20$ / $-10$ \\
\bottomrule
\end{tabular}
\end{table}

\begin{table*}[ht!]
\centering
\small
\caption{Dataset summary across three evaluation domains.}
\label{tab:datasets}
\resizebox{\textwidth}{!}{%
\begin{tabular}{lllcccclll}
\toprule
\textbf{Dataset} & \textbf{Domain} & \textbf{Source} & \textbf{Size} & \textbf{Entities} & \textbf{Queries} & \textbf{Search} & \textbf{Embedding} & \textbf{Chunking} & \textbf{Labeling} \\
\midrule
Seller Forums & E-comm. seller & Public & ${\sim}$8K & 6 & 36 & Hybrid & Titan V2~\citep{aws2024titanembedv2} (1024d) & 300 tok / 20\% overlap & LLM-enriched metadata \\
Yelp Hotels & Hospitality & Yelp Open & 14K & 10 & 60 & Semantic & MiniLM-L6 (384d) & Whole review & LLM-extracted (Claude) \\
OpinRank Cars & Automotive & OpinRank & 13K & 10 & 60 & Semantic & MiniLM-L6 (384d) & Whole review & LLM-extracted \\
\bottomrule
\end{tabular}%
}
\end{table*}

\paragraph{Entity Selection.} Three stages. First, LLM-based entity extraction with domain-specific seed lists. Then diversity filtering: Shannon entropy $H \geq 0.6$, minority sentiment $\geq 10\%$, mention count $\geq 100$. Finally, top-$n$ selection ranked by entropy descending.

\subsection{Query Templates}
\label{app:queries}

Table~\ref{tab:query-templates} lists the exact query templates used to generate evaluation questions. Each entity is instantiated into all 6 templates, yielding 156 total queries (36 Seller Forums + 60 Yelp + 60 OpinRank).

\begin{table*}[ht!]
\centering
\small
\caption{Query templates per domain. Each template is instantiated with every selected entity (6 for Seller Forums, 10 for Yelp/OpinRank), producing 6 questions per entity (2 breadth, 2 polar, 2 segment).}
\label{tab:query-templates}
\resizebox{\textwidth}{!}{%
\begin{tabular}{llp{4.2cm}p{4.2cm}p{4.2cm}}
\toprule
\textbf{Type} & \textbf{Idx} & \textbf{Seller Forums} & \textbf{Yelp Hotels} & \textbf{OpinRank Cars} \\
\midrule
breadth & 0 & What do sellers think about \{entity\}? & What do guests think about \{entity\}? & What do owners think about \{entity\}? \\
breadth & 1 & Summarize seller opinions on \{entity\}. & Summarize guest opinions on \{entity\}. & Summarize owner opinions on \{entity\}. \\
\midrule
polar & 0 & What are the biggest complaints about \{entity\}? & What are the biggest complaints about \{entity\}? & What are the biggest complaints about \{entity\}? \\
polar & 1 & What positive experiences have sellers had with \{entity\}? & What positive experiences have guests had with \{entity\}? & What do owners praise most about \{entity\}? \\
\midrule
segment & 0 & How do small sellers vs large sellers feel about \{entity\}? & How do business travelers vs leisure travelers feel about \{entity\}? & How do commuters vs enthusiasts feel about \{entity\}? \\
segment & 1 & Do new sellers and experienced sellers differ in their views of \{entity\}? & Do solo guests and families differ in their views of \{entity\}? & Do new buyers and long-term owners differ in their views of \{entity\}? \\
\bottomrule
\end{tabular}%
}
\end{table*}

\paragraph{Entities.} \textbf{Seller Forums} (6): A+ Content, Account Health Rating, Brand Stores, Community Management, Coupons, Seller Education. \textbf{Yelp Hotels} (10): Booking Experience, Breakfast, Breakfast Quality, Business Center, Casino, Loyalty Program, Nightly Rate, Pool, Room Quality, Shuttle Service. \textbf{OpinRank Cars} (10): Oil Change Interval, Dealer Experience, Vehicle Size, Sales Experience, Fuel Tank Size, Resale Value, City MPG, Remote Start, Climate Control, Ground Clearance.

\subsection{Prompt Library}
\label{app:prompts}

All prompts used in the WARP pipeline. Template variables shown in \texttt{\{braces\}}.

\subsubsection{Opinion Extraction}
\label{app:prompts:extraction}

\promptheading{Registry-Guided Extraction (Open-Domain)}{A single LLM call per document extracts entities and sentiment-intensity labels using a domain-specific tiered seed list while remaining open to discovering new entities.}
\label{prompt:extraction}
\begin{promptcontent}
\textbf{System:} You extract structured opinion data from reviews. Output ONLY valid JSON arrays. No markdown, no explanation.\medskip

\textbf{User:} Extract entity-level opinions from this review.\medskip

\texttt{[\{registry\_block\} --- domain-specific tiered entity list injected here]}\medskip

\textbf{Instructions:}\\
1. Identify ALL entities/aspects the reviewer discusses.\\
2. If an entity matches the registry above, use the EXACT registry name. Prefer the most specific tier (Tier 3 > Tier 2 > Tier 1).\\
3. If the reviewer discusses something NOT in the registry, create a short descriptive entity name (open discovery).\\
4. For EACH entity mentioned, output:\\
\quad- entity: the entity name (registry canonical or new)\\
\quad- sentiment: ``positive'', ``negative'', ``neutral'', or ``mixed''\\
\quad- intensity: ``high'', ``medium'', or ``low''\\
\quad- evidence: exact quote from the review supporting this opinion (max 120 chars)\medskip

Only include entities explicitly discussed. Return \texttt{[]} if none found.\\
Output ONLY a JSON array.\medskip

Review:\\
\texttt{\{review\_text\}}
\end{promptcontent}

\noindent For closed-domain settings (e.g., Yelp Hotels), the extraction prompt uses a fixed aspect list with keyword mappings instead of the tiered registry, but the output schema is identical.

\subsubsection{Answer Generation}
\label{app:prompts:generation}

\promptheading{Opinion Summary Generator}{Instructs the LLM to synthesize a proportional opinion summary from retrieved documents.}
\label{prompt:generator}
\begin{promptcontent}
\textbf{System:} You are summarizing community opinions from review data. Use ONLY the provided documents. Represent opinions proportionally --- include minority views, not just the majority. Be specific and cite reviewer experiences where possible.\medskip

\textbf{User:} Question: \texttt{\{question\}}\medskip

Reviews:\\
\texttt{\{docs\}}\medskip

Summarize what the community thinks. Be thorough --- cover the full range of opinions proportionally.\\
Do not add opinions beyond what's in the documents.
\end{promptcontent}

\subsubsection{Pairwise Generation Evaluation (Judge)}
\label{app:prompts:judge}

Each answer pair is evaluated via position-controlled blind pairwise comparison (positions swapped in the second call for debiasing). The judge scores three dimensions; the \textsc{opinion fairness} dimension is swapped between variants in our ablation (\S\ref{app:prompt-ablation}).

\promptheading{Pairwise Judge Template}{Full evaluation template: blind comparison on opinion fairness, informativeness, and groundedness.}
\label{prompt:judge}
\begin{promptcontent}
\textbf{System:} You are an expert evaluator comparing two answers that summarize community opinions. You do NOT know which retrieval method produced which answer. Judge strictly on the quality of the answers themselves.\smallskip

\textbf{User:} You are evaluating two answers to the same question about community opinions. Both answers were generated from different sets of reviews retrieved for the same question.\smallskip

Question: \texttt{\{question\}}\smallskip

--- Answer A ---\\
\texttt{\{answer\_a\}}\smallskip

--- Answer B ---\\
\texttt{\{answer\_b\}}\smallskip

Compare the answers on these dimensions. For each, pick ``A'', ``B'', or ``tie'':\smallskip

1. OPINION FAIRNESS: \texttt{[fairness variant inserted --- see Prompts~\ref{prompt:fair-propA} and~\ref{prompt:fair-propB} below]}\smallskip

2. INFORMATIVENESS: Which answer is more helpful, specific, and actionable for someone trying to understand what the community thinks? Consider breadth of points covered, specificity, and usefulness.\smallskip

3. GROUNDEDNESS: Which answer appears more grounded in actual community member experiences? Look for attribution to specific perspectives vs vague generalizations that could be hallucinated.\smallskip

Respond ONLY with JSON:\\
\texttt{\{"fairness": "A"|"B"|"tie", "informativeness": "A"|"B"|"tie", "groundedness": "A"|"B"|"tie", "reasoning": "<1-2 sentences>"\}}
\end{promptcontent}

\noindent The baseline evaluation uses a generic fairness instruction (``which answer more proportionally represents the full range of community opinions, including minority views?''). Our ablation replaces it with the two variants below.

\promptheading{Fairness Variant A --- Proportional Accuracy}{Provides ground-truth $P_{\text{pop}}$ to the judge; penalizes over-representing minority opinions.}
\label{prompt:fair-propA}
\begin{promptcontent}
OPINION FAIRNESS (Proportional Accuracy): The actual community opinion distribution for this topic is: \texttt{\{p\_pop\_description\}}. Which answer more accurately reflects these real proportions? The ideal answer should make the reader walk away with a correct sense of how common each view is: a dominant opinion should dominate the summary, a rare opinion should be mentioned but not over-emphasized. An answer that gives equal weight to a 10\% minority and a 70\% majority is MISLEADING, even if well-intentioned. Pick the answer whose emphasis better matches the actual distribution above.
\end{promptcontent}

\noindent Where \texttt{\{p\_pop\_description\}} is dynamically filled per entity from the ground-truth $P_{\text{pop}}$, e.g.: ``approximately 65\% positive (mostly satisfied), 25\% negative (dissatisfied), 10\% neutral.''

\promptheading{Fairness Variant B --- Viewpoint Coverage}{Rewards distinct viewpoint coverage; favors surfacing rare perspectives even if disproportionate.}
\label{prompt:fair-propB}
\begin{promptcontent}
OPINION FAIRNESS (Viewpoint Coverage): Which answer covers MORE DISTINCT VIEWPOINTS from the community, giving voice to all perspectives including rare and minority ones? An answer that only represents the majority view --- even if that majority is large --- is LESS fair than one that surfaces unique perspectives readers might not otherwise encounter. The ideal answer ensures no viewpoint goes unheard, even if that means giving disproportionate space to rare opinions.
\end{promptcontent}

\subsubsection{Retrieval Augmentation}
\label{app:prompts:retrieval-aug}

\promptheading{Pole-Biased Multi-Query Templates}{During pool expansion (Stage~1.5), three sentiment-pole queries recover documents that cosine retrieval buries.}
\label{prompt:pole}
\begin{promptcontent}
\textbf{Positive:} What positive experiences have sellers had with \texttt{\{entity\}}?\medskip

\textbf{Negative:} What are the biggest complaints about \texttt{\{entity\}}?\medskip

\textbf{Neutral:} What factual information do sellers share about \texttt{\{entity\}}?
\end{promptcontent}

\noindent Evaluation queries (6 per entity) are generated from three templates---breadth (``What do \texttt{\{persona\}} think about \texttt{\{entity\}}?''), polar (``What are the biggest complaints...''), and segment (``How do \texttt{\{segment\_A\}} vs \texttt{\{segment\_B\}} feel...'')---with domain-adapted persona and segment terms (e.g., ``sellers'' / ``small sellers vs large sellers'').

\section{Algorithm Details}
\label{app:algorithms}
\label{app:pool-algorithms}

Table~\ref{tab:notation} summarizes the key symbols used throughout.

\begin{table}[ht!]
\centering
\small
\caption{Notation reference.}
\label{tab:notation}
\begin{tabular}{ll}
\toprule
\textbf{Symbol} & \textbf{Meaning} \\
\midrule
$d$ & A candidate document \\
$e$ & Target entity \\
$k$ & Output size (documents returned) \\
$N$ & Candidate pool size \\
$SI(d)$ & Sentiment-intensity label of $d$ \\
$P_{\text{pop}}$ & Ground-truth population distribution \\
$P_{\text{pool}}$ & Pool-level empirical distribution \\
$W_1(P, Q)$ & Wasserstein-1 distance between $P$ and $Q$ \\
$\lambda$ & Relevance--calibration trade-off weight \\
$\delta$ & Deficit threshold for pool expansion \\
$n_{\text{exp}}$ & Budget (docs per deficit pole) \\
$\text{rel}(d)$ & Relevance score of document $d$ \\
$\text{EM}$ & Entity-matched subset of pool \\
\bottomrule
\end{tabular}
\end{table}

\subsection{Entity-Gated Re-retrieval}

The two-pass entity-gated pool construction recovers entity-matched documents that pure relevance ranking buries.

\begin{algorithm}[H]
\caption{Entity-Gated Re-retrieval}
\begin{algorithmic}[1]
\REQUIRE Results $R$ (by score), entity $e$, sizes $n_1, n_2$, threshold $\tau$
\STATE $P_1 \leftarrow R[1 : n_1]$ \hfill (top-$n_1$ by relevance)
\STATE $C_2 \leftarrow \{d \in R \setminus P_1 : \text{entity}(d) = e \wedge \text{score}(d) \geq \tau\}$
\STATE Sort $C_2$ by $|SI(d)|$ descending
\STATE $P_2 \leftarrow C_2[1 : n_2]$
\STATE Pool $\leftarrow \text{dedup}(P_1 \cup P_2)$, sorted by score
\RETURN Pool
\end{algorithmic}
\end{algorithm}

\subsection{Adaptive Distribution-Aware Pool Expansion}

When entity-matched documents under-represent a sentiment pole relative to $P_{\text{pop}}$, pole-biased queries (\S\ref{app:prompts:retrieval-aug}) retrieve additional candidates before re-ranking. The cold-start path ($\text{EM} = \emptyset$) treats every pole with non-negligible mass as a deficit, ensuring new entities still receive balanced pools. In practice, one to three deficit poles trigger per entity; $n_{\text{exp}}$ caps total expansion cost at a fixed multiple of $k$.

\begin{algorithm}[H]
\caption{Adaptive Pool Expansion}
\label{alg:adaptive-expansion}
\begin{algorithmic}[1]
\REQUIRE Pool $C_0$, entity $e$, $P_{\text{pop}}$, threshold $\delta$, budget $n_{\text{exp}}$
\STATE $\text{EM} \leftarrow \{d \in C_0 : \text{entity}(d) = e\}$
\IF{$\text{EM} = \emptyset$}
    \STATE DeficitPoles $\leftarrow \{s : P_{\text{pop}}(s) > 0.01\}$
\ELSE
    \STATE $P_{\text{pool}}(s) \leftarrow |\{d \in \text{EM} : SI(d) = s\}| / |\text{EM}|$
    \STATE DeficitPoles $\leftarrow \{s : P_{\text{pool}}(s) < \delta \cdot P_{\text{pop}}(s)\}$
\ENDIF
\IF{DeficitPoles $= \emptyset$}
    \RETURN $C_0$
\ENDIF
\STATE Retrieve $n_{\text{exp}}$ entity-matched docs per deficit pole
\STATE Pool $\leftarrow \text{dedup}(C_0 \cup C_{\text{exp}})$
\RETURN Pool
\end{algorithmic}
\end{algorithm}

\subsection{$W_1$-MMR}
\label{app:w1mmr}

The $W_1$-MMR variant blends relevance with marginal Wasserstein reduction, mirroring the MMR trade-off while targeting distributional calibration rather than pairwise dissimilarity.

\begin{algorithm}[H]
\caption{$W_1$-MMR}
\begin{algorithmic}[1]
\REQUIRE Pool $C$, scores $\text{rel}(\cdot)$, target $P_{\text{pop}}$, $\lambda$, size $k$
\STATE $S \leftarrow \emptyset$
\WHILE{$|S| < k$}
    \STATE $W_1^{\text{cur}} \leftarrow \begin{cases} 60 & \text{if } S = \emptyset \\ W_1(\text{EmpDist}(S), P_{\text{pop}}) & \text{else} \end{cases}$
    \FOR{each $d \in C \setminus S$}
        \STATE $\Delta W_1 \leftarrow W_1^{\text{cur}} - W_1(\text{EmpDist}(S \cup \{d\}), P_{\text{pop}})$
        \STATE $\text{score}(d) \leftarrow \lambda \cdot \text{rel}(d) + (1{-}\lambda) \cdot \Delta W_1 / W_1^{\text{cur}}$
    \ENDFOR
    \STATE $d^* \leftarrow \arg\max_{d \in C \setminus S} \text{score}(d)$
    \STATE $S \leftarrow S \cup \{d^*\}$
\ENDWHILE
\RETURN $S$
\end{algorithmic}
\end{algorithm}

\subsection{Label Acquisition}
\label{app:labeling}

Per-document SI labels come from a single LLM pass at indexing time ($<$200\,ms/doc). Labeling is cheap. The Seller Forums corpus (${\sim}$8K documents) labels in under 4 minutes with 150 concurrent workers. $P_{\text{pop}}$ is just a count aggregation per entity ($O(n)$, milliseconds); incremental updates on ingestion avoid recomputation entirely.

\subsection{End-to-End Latency Breakdown}
\label{app:latency}

\textbf{Hardware.} Intel Xeon Platinum 8175M @ 2.50\,GHz, 8 cores (single-threaded execution), no GPU.

WARP's re-ranking step runs entirely in NumPy on the retrieved candidate distribution --- no additional model calls. End-to-end latency therefore decomposes into three components (Table~\ref{tab:latency-breakdown}):

\begin{table*}[ht!]
\centering
\small
\caption{Per-component latency breakdown. Re-ranking is hardware-only (no model inference); retrieval and re-retrieval depend on the vector-store backend. p99 for re-ranking; typical/median for the rest.}
\label{tab:latency-breakdown}
\begin{tabular}{lcc}
\toprule
\textbf{Component} & \textbf{Local FAISS} & \textbf{Managed Bedrock Knowledge Base} \\
\midrule
Retrieval                              & $<$50\,ms   & ${\sim}$1.5\,s \\
Re-retrieval$^\dagger$                 & $<$5\,ms    & ${\sim}$100\,ms \\
Re-ranking (p99)                       & $<$310\,ms  & $<$310\,ms \\
\midrule
\textbf{End-to-end}                    & \textbf{$<$365\,ms} & \textbf{${\sim}$1.9\,s} \\
\bottomrule
\multicolumn{3}{l}{\footnotesize $^\dagger$ Only fires when pool expansion (Entity-Gated / Adaptive) detects a deficit.} \\
\end{tabular}
\end{table*}

Re-ranking latency is dominated by CDF differencing over $m{=}7$ bins and is backend-independent. Retrieval and re-retrieval dominate managed-endpoint deployments; for local FAISS the whole pipeline fits inside a 500\,ms sub-second SLA.

\begin{table*}[ht!]
\centering
\small
\caption{Per-query re-ranking latency percentiles ($N{=}96$ queries: 36 Seller Forums + 60 Yelp Hotels, single-threaded on the hardware above). p99 drives production SLAs; max reported for tail auditing. All WARP variants stay under 310\,ms at p99.}
\label{tab:latency-percentiles}
\begin{tabular}{ll r rrrr}
\toprule
\textbf{Domain} & \textbf{Method} & \textbf{Queries} & \textbf{p50 (ms)} & \textbf{p95 (ms)} & \textbf{p99 (ms)} & \textbf{Max (ms)} \\
\midrule
Seller Forums & $W_1$ Minimizer & 36 & \textbf{0.1}   & \textbf{38.7}  & \textbf{45.5}  & \textbf{48.0}  \\
Seller Forums & $W_1$-MMR       & 36 & 156.8 & 163.1 & 166.2 & 167.4 \\
Seller Forums & EG + $W_1$ Min  & 36 & 0.6   & 211.0 & 216.2 & 218.8 \\
Seller Forums & DPP             & 36 & 47.8  & 49.1  & 51.9  & 53.0  \\
\midrule
Yelp Hotels   & $W_1$ Minimizer & 60 & \textbf{65.0}  & 270.8 & 289.9 & 295.3 \\
Yelp Hotels   & $W_1$-MMR       & 60 & 81.6  & 284.7 & 301.6 & 303.5 \\
Yelp Hotels   & EG + $W_1$ Min  & 60 & 145.7 & 302.0 & 307.6 & 312.5 \\
Yelp Hotels   & DPP             & 60 & 103.5 & \textbf{127.4} & \textbf{129.3} & \textbf{131.3} \\
\bottomrule
\end{tabular}
\end{table*}

The $W_1$ Minimizer's near-zero p50 on Seller Forums (0.1\,ms) reflects the sparse entity-filtered pool: few candidates match the queried entity, so the greedy loop terminates almost immediately. p99 stays under 310\,ms across all WARP variants on both domains, and end-to-end with local FAISS retrieval fits in ${<}365$\,ms --- comfortably inside a sub-second SLA. Managed vector stores add ${\sim}$1.5\,s of retrieval latency that WARP does not remove.

\section{Retrieval Ablations}
\label{app:sec:ablations}

We ablate key components of the retrieval pipeline: distance metric choice, entity filtering, pool expansion, convergence, per-entity variation, and sensitivity to hyperparameters.

\subsection{JS vs.\ $W_1$: Metric Comparison and Entity-Match Confound}
\label{app:kl-ablation}
\label{app:em-confound}

This section unpacks the JS-vs-$W_1$ comparison summarized in Section~\ref{sec:ordinal}. We rerun the four calibrated methods (KL(JS) Minimizer, $W_1$ Minimizer, KL(JS)-MMR, $W_1$-MMR) on Yelp and OpinRank at $k{=}20$, holding the greedy structure and entity filtering fixed so only the distance metric varies within each row-pair.

Table~\ref{tab:em-confound} shows KL(JS)-MMR beating $W_1$-MMR on Yelp and OpinRank $W_1$. On its face this reverses the ordinal-metric advantage the Minimizer comparison establishes. It doesn't: the two hybrids retrieve very different candidate pools, and the metric is being scored on top of that difference.

\begin{table}[ht!]
\centering
\small
\caption{Entity-match confound in hybrid variants. KL(JS)-MMR retrieves $\sim$1.7--2$\times$ more entity-matched documents than $W_1$-MMR on both Yelp and OpinRank. In the controlled Minimizer comparison (same entity filter, EM equalized) $W_1$ wins on both domains --- the KL(JS)-MMR $W_1$ advantage in the uncontrolled comparison traces to the EM gap, not to metric superiority.}
\label{tab:em-confound}
\begin{tabular}{l cc cc}
\toprule
& \multicolumn{2}{c}{\textbf{Yelp}} & \multicolumn{2}{c}{\textbf{OpinRank}} \\
\cmidrule(lr){2-3} \cmidrule(lr){4-5}
\textbf{Method} & $W_1$ & EM\% & $W_1$ & EM\% \\
\midrule
KL (JS)-MMR       & \textbf{2.51} & 38.2\%          & \textbf{3.32} & 36.1\% \\
$W_1$-MMR         & 4.02          & 19.4\%          & 3.41          & 21.8\% \\
\midrule
KL (JS) Minimizer & 3.00          & 93.8\%          & 4.54          & 85.8\% \\
$W_1$ Minimizer   & \textbf{2.76} & 93.8\%          & \textbf{4.14} & 85.8\% \\
\bottomrule
\end{tabular}
\end{table}

In the top block (hybrids, uncontrolled EM), KL(JS)-MMR retrieves nearly twice as many entity-matched documents as $W_1$-MMR (38.2\% vs.\ 19.4\% on Yelp; 36.1\% vs.\ 21.8\% on OpinRank), and its lower $W_1$ tracks that EM gap rather than the choice of ordinal-vs-categorical ground cost. In the bottom block (Minimizers, controlled EM), both algorithms filter to the same entity-matched pool (EM${=}93.8\%$ Yelp, $85.8\%$ OpinRank), and $W_1$ wins on both domains --- 2.76 vs.\ 3.00 on Yelp, 4.14 vs.\ 4.54 on OpinRank. The controlled comparison is where the ordinal metric earns its billing; the uncontrolled hybrid comparison is a retrieval-pool artifact.

\subsection{Entity-Filtered Control (Yelp)}
\label{app:entity-control}

\begin{table}[ht!]
\centering
\small
\caption{Entity-filter control experiment (Yelp Hotels, $k{=}20$). Entity filtering alone reduces $W_1$ by 45.7\%; $W_1$ Minimizer achieves 62.6\% additional reduction beyond the entity-filter control.}
\resizebox{\columnwidth}{!}{%
\begin{tabular}{lcccc}
\toprule
\textbf{Approach} & \textbf{EM\%} & $W_1$ & \textbf{$\Delta$ vs Top-K} & \textbf{$\Delta$ vs EF+Top-K} \\
\midrule
Top K (no filter) & 40.1 & 13.60 & --- & --- \\
EntityFilter + Top-K & 100.0 & 7.38 & $-45.7\%$ & --- \\
EntityFilter + MMR & 100.0 & 6.99 & $-48.6\%$ & $-5.3\%$ \\
EntityFilter + DPP & 100.0 & 5.64 & $-58.5\%$ & $-23.6\%$ \\
$W_1$ Minimizer & 93.8 & 2.76 & $-79.7\%$ & $-62.6\%$ \\
$W_1$-MMR & 19.4 & 4.02 & $-70.4\%$ & $-45.5\%$ \\
EG + $W_1$ Min & 99.2 & 1.52 & $-88.8\%$ & $-79.4\%$ \\
AE + $W_1$ Min & 95.9 & 1.66 & $-87.8\%$ & $-77.5\%$ \\
\bottomrule
\end{tabular}%
}
\end{table}

On sparse Seller Forums (Top-$k$ baseline EM${\sim}$40\%; entity-filter ceiling ${\sim}$51\%), entity filtering provides only $-11.6\%$ $W_1$ reduction (vs.\ $-45.7\%$ on Yelp, where the entity filter reaches 100\% EM), confirming that $W_1$-MMR's full-pool scoring drives gains in entity-sparse domains ($-37.7\%$ beyond the entity-filter baseline).

\subsection{Adaptive-$k$ Convergence}
\label{app:adaptive-k}

\begin{figure*}[ht!]
\centering
\includegraphics[width=0.92\textwidth]{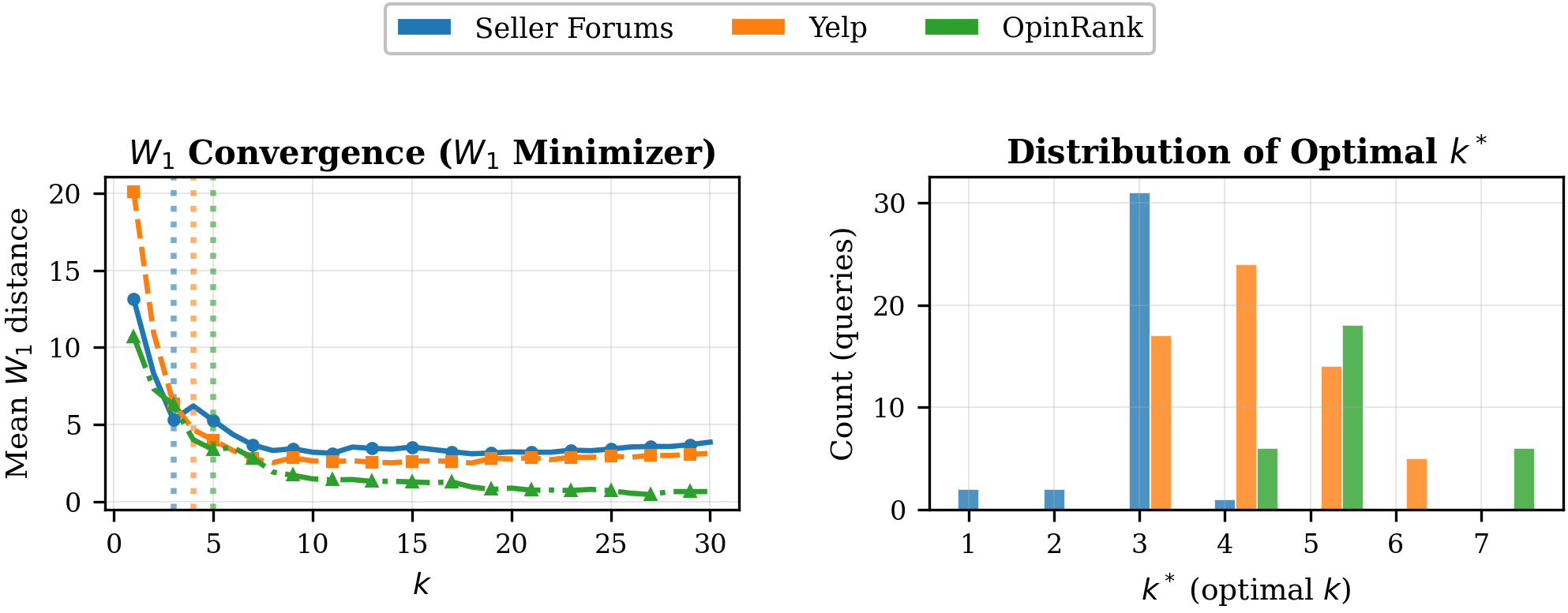}
\caption{$W_1$ Minimizer convergence vs.\ output size $k$. Median convergence point $k^*$ (where $\Delta W_1 < 0.5$ per additional document): Seller Forums $k^*{=}3$, Yelp $k^*{=}4$, OpinRank $k^*{=}5$.}
\label{fig:adaptive-k}
\end{figure*}

$W_1$ Minimizer hits diminishing returns at $k^*{=}3$ (Seller Forums), $k^*{=}4$ (Yelp), $k^*{=}5$ (OpinRank) (Figure~\ref{fig:adaptive-k}). An adaptive-$k$ strategy could therefore feed far fewer documents to the LLM without losing distributional fidelity. This also implies robustness to imprecise $P_{\text{pop}}$ estimates.

\subsection{$W_2$ vs.\ $W_1$ Comparison}
\label{app:w2}

A natural concern is whether optimizing $W_1$ (linear ground cost) sacrifices performance under higher-order transport metrics. We compute $W_2$ (quadratic ground cost) for all methods on Yelp Hotels ($k{=}20$). Across the distributions produced by these methods, we observe that the method ranking under $W_1$ is preserved under $W_2$, with a linear relationship (slope $= 1.19$, $R^2 > 0.99$). $W_1$ is chosen for its $O(m)$ closed-form computation.

\subsection{Per-Entity Breakdowns}
\label{app:per-entity}

Aggregate results in Table~\ref{tab:main} mask substantial per-entity variation. The gains concentrate on entities whose baseline distributions are most skewed, precisely where minority-opinion misrepresentation does the most damage. We break out per-entity $W_1$ for all three domains below.

\subsubsection{Yelp Hotels ($k{=}20$)}

$W_1$ Minimizer lands below 2.0 on 7 of 10 entities. That is near-perfect calibration. Three outliers remain above 2.0: Business Center (9.76), Shuttle Service (5.78), and Loyalty Program (4.07), all corresponding to entities with extreme distributional skew and thin pool diversity. Even greedy optimization cannot fully close the gap within $k{=}20$ selections under those conditions.

\begin{table}[ht!]
\centering
\small
\caption{Per-entity breakdown (Yelp Hotels, 10 aspects). $W_1$ Minimizer achieves $<$2.0 on 7/10 entities.}
\resizebox{\columnwidth}{!}{%
\begin{tabular}{l cccc}
\toprule
\textbf{Entity} & \textbf{Top-$k$ $W_1$} & $W_1$ \textbf{Min} & $W_1$\textbf{-MMR} & \textbf{WassRank OT} \\
\midrule
Booking Experience & 18.30 & 1.95 & 3.78 & 1.58 \\
Breakfast & 7.55 & 1.48 & 2.87 & 1.42 \\
Breakfast Quality & 5.78 & 0.87 & 3.79 & 0.65 \\
Business Center & 24.51 & 9.76 & 6.72 & 9.26 \\
Casino & 14.79 & 0.86 & 4.07 & 0.62 \\
Loyalty Program & 16.31 & 4.07 & 4.49 & 2.95 \\
Nightly Rate & 7.49 & 0.82 & 3.45 & 0.66 \\
Pool & 10.79 & 0.73 & 2.68 & 0.46 \\
Room Quality & 8.55 & 1.24 & 2.67 & 2.18 \\
Shuttle Service & 21.95 & 5.78 & 5.70 & 5.38 \\
\bottomrule
\end{tabular}%
}
\end{table}

\subsubsection{OpinRank Cars ($k{=}20$, top 10 by entropy)}

OpinRank entities vary widely in mention count (39--1,465) and minority fraction (19--45\%). High-count entities with moderate minority shares calibrate well: City MPG, for instance, has 1,465 mentions at 25\% minority and reaches $W_1{=}0.38$. Hard cases persist. Vehicle Size has only 42 mentions; Dealer Experience starts from a $W_1{=}23.07$ baseline. Both resist full correction.

\begin{table*}[ht!]
\centering
\small
\caption{Per-entity breakdown (OpinRank, LLM-extracted entities). Count = opinion mentions, H = Shannon entropy, Min\% = minority sentiment fraction.}
\begin{tabular}{l ccc cccc}
\toprule
\textbf{Entity} & \textbf{Count} & \textbf{H} & \textbf{Min\%} & \textbf{Top-$k$} & $W_1$ \textbf{Min} & $W_1$\textbf{-MMR} & \textbf{WassRank OT} \\
\midrule
Oil Change Interval & 144 & 1.49 & 21\% & 3.96 & 1.08 & 2.95 & 0.78 \\
Dealer Experience & 425 & 1.47 & 30\% & 23.07 & 11.71 & 3.40 & 10.28 \\
Vehicle Size & 42 & 1.37 & 19\% & 0.00 & 8.31 & 6.19 & 8.31 \\
Sales Experience & 776 & 1.33 & 28\% & 15.66 & 3.01 & 3.19 & 2.57 \\
Fuel Tank Size & 269 & 1.32 & 36\% & 13.67 & 3.06 & 3.25 & 3.05 \\
Resale Value & 427 & 1.30 & 39\% & 12.19 & 2.24 & 2.40 & 2.33 \\
City MPG & 1465 & 1.29 & 25\% & 9.12 & 0.38 & 2.78 & 0.38 \\
Remote Start & 39 & 1.25 & 41\% & 8.80 & 4.52 & 3.54 & 4.52 \\
Climate Control & 727 & 1.24 & 45\% & 9.95 & 0.95 & 2.43 & 0.94 \\
Ground Clearance & 201 & 1.22 & 33\% & 23.55 & 6.12 & 4.00 & 5.50 \\
\bottomrule
\end{tabular}
\end{table*}

\section{Statistical Validity}
\label{app:sec:stat-validity}

Two orthogonal concerns for the significance testing: multiple comparisons across many method $\times$ domain combinations, and non-independent samples within a domain (queries drawn from the same entity share an underlying opinion distribution). We address both.

\subsection{Benjamini--Hochberg Correction}
\label{app:bh}

We ran 57 paired Wilcoxon tests spanning three domains, multiple re-ranking variants (baselines and $W_1$ family), and all evaluated $k$ values. Benjamini--Hochberg FDR correction at $\alpha{=}0.05$ leaves 39/57 tests surviving. Table~\ref{tab:bh} shows the 14 $W_1$-family tests: every one survives on every domain. The 18 non-surviving tests come from baseline re-ranking methods outside our contribution set.

\begin{table}[ht!]
\centering
\small
\caption{Benjamini--Hochberg FDR correction ($\alpha{=}0.05$) on paired-Wilcoxon $p$-values for the $W_1$ family. Every $W_1$-family algorithm survives on every domain (14/14). Reduction is per-query paired reduction relative to Top-$k$; aggregate reductions in Table~\ref{tab:main} weight queries differently. $^{***}p_{\text{adj}}{<}10^{-3}$, $^{**}p_{\text{adj}}{<}10^{-2}$, $^{*}p_{\text{adj}}{<}0.05$.}
\label{tab:bh}
\resizebox{\columnwidth}{!}{%
\begin{tabular}{llrrl}
\toprule
\textbf{Domain} & \textbf{Approach} & $N$ & \textbf{Reduction} & $p_{\text{adj}}$ \\
\midrule
Yelp          & $W_1$ Minimizer$^{***}$  & 57 & $+81.4\%$ & $<$0.000001 \\
Yelp          & $W_1$-MMR$^{***}$        & 57 & $+70.2\%$ & $<$0.000001 \\
Yelp          & EG + $W_1$ Min$^{***}$   & 57 & $+89.8\%$ & $<$0.000001 \\
Yelp          & AE + $W_1$ Min$^{***}$   & 57 & $+89.1\%$ & $<$0.000001 \\
Yelp          & WassRank OT$^{***}$      & 57 & $+83.3\%$ & $<$0.000001 \\
OpinRank      & $W_1$ Minimizer$^{***}$  & 54 & $+72.4\%$ & $<$0.000001 \\
OpinRank      & $W_1$-MMR$^{***}$        & 54 & $+76.7\%$ & $<$0.000001 \\
OpinRank      & EG + $W_1$ Min$^{***}$   & 54 & $+72.4\%$ & $<$0.000001 \\
OpinRank      & AE + $W_1$ Min$^{***}$   & 54 & $+72.4\%$ & $<$0.000001 \\
OpinRank      & WassRank OT$^{***}$      & 54 & $+74.7\%$ & $<$0.000001 \\
Seller Forums & $W_1$ Minimizer$^{***}$  & 24 & $+59.1\%$ & 0.000427 \\
Seller Forums & $W_1$-MMR$^{**}$         & 24 & $+20.4\%$ & 0.004552 \\
Seller Forums & EG + $W_1$ Min$^{*}$     & 24 & $+53.7\%$ & 0.049 \\
Seller Forums & AE + $W_1$ Min$^{*}$     & 24 & $+53.7\%$ & 0.049 \\
\bottomrule
\end{tabular}%
}
\end{table}

\subsection{Entity-Clustered Bootstrap}
\label{app:bootstrap}

Paired-Wilcoxon assumes independent paired observations. Queries drawn from the same entity are not independent, since they share the same underlying opinion distribution. We therefore resample at the entity level: 10{,}000 bootstrap iterations, drawing entities with replacement and recomputing the mean per-query $W_1$ improvement. All 95\% CIs exclude zero for every $W_1$ method on every domain (Table~\ref{tab:bootstrap}).

\begin{table}[ht!]
\centering
\small
\caption{Entity-clustered bootstrap (10{,}000 resamples). Mean improvement in $W_1$ relative to Top-$k$; 95\% CIs computed by resampling entities with replacement. All CIs exclude zero.}
\label{tab:bootstrap}
\resizebox{\columnwidth}{!}{%
\begin{tabular}{llccc}
\toprule
\textbf{Dataset} & \textbf{Method} & \textbf{\#Entities} & \textbf{Mean $\Delta W_1$} & \textbf{95\% CI} \\
\midrule
Yelp     & $W_1$ Minimizer & 10 & 10.85 & $[8.26, 13.37]$ \\
Yelp     & EG + $W_1$ Min  & 10 & 12.09 & $[8.72, 15.37]$ \\
Yelp     & $W_1$-MMR       & 10 &  9.58 & $[6.40, 12.80]$ \\
OpinRank & $W_1$ Minimizer &  9 &  9.65 & $[6.92, 12.34]$ \\
OpinRank & $W_1$-MMR       &  9 & 10.22 & $[6.48, 14.17]$ \\
Seller Forums   & $W_1$ Minimizer &  4 &  5.32 & $[0.67, 10.38]$ \\
Seller Forums   & $W_1$-MMR       &  4 &  1.84 & $[0.49,  3.25]$ \\
\bottomrule
\end{tabular}%
}
\end{table}

Seller Forums' wider CIs (e.g., $W_1$ Minimizer $[0.67, 10.38]$) reflect its small entity count ($N{=}4$), not a qualitatively different effect --- the lower bound still exceeds zero. The aggregate signal is cross-domain consistency across 10 Yelp, 9 OpinRank, and 4 Seller Forums entities.

\section{Independent-Target Validation}
\label{app:star-validation}

Our LLM-derived sentiment-intensity labels define both the population target $P_{\text{pop}}$ used at inference and the $W_1$ evaluation metric applied to the retrieved evidence. If the labeler had systematic biases, calibrating and measuring against it could inflate the numbers. We cross-validate against Yelp's native 1--5 star ratings, which are ordinal, human-authored at review time, and participate in \emph{no part} of the WARP pipeline --- neither retrieval, nor re-ranking, nor the $P_{\text{pop}}$ used at inference. Section~\ref{sec:star-validation} in the main body summarizes; this appendix reports the underlying numbers.

A high correlation on individual labels is a necessary but not sufficient condition: the entity-level $P_{\text{pop}}$ is an aggregate, and a labeler could still consistently mis-estimate it. We therefore recompute $P_{\text{pop}}$ from star ratings alone for each of the 10 Yelp entities and quantify the divergence from the LLM-derived $P_{\text{pop}}$ actually used at inference (Table~\ref{tab:star-ppop}). Mean $W_1$ divergence: 2.88. Under a triangle-inequality bound, WARP's retrievals stay within $W_1{=}5.64$ of the star-derived target while Top-$k$ is at least $W_1{=}10.72$ away --- a worst-case reduction of at least 47.4\% against a target the labeler never touched.

\begin{table}[ht!]
\centering
\small
\caption{Star-derived vs.\ LLM-derived $P_{\text{pop}}$ per Yelp entity. Mean $W_1$ gap: 2.88. Bounds: $W_1(\text{WARP}, \text{star-}P_{\text{pop}}) \leq 5.64$, $W_1(\text{Top-}k, \text{star-}P_{\text{pop}}) \geq 10.72$; worst-case reduction $\geq 47.4\%$.}
\label{tab:star-ppop}
\resizebox{\columnwidth}{!}{%
\begin{tabular}{lrc}
\toprule
\textbf{Entity} & \textbf{$N$ reviews} & $W_1(\text{star-}P_{\text{pop}}, \text{LLM-}P_{\text{pop}})$ \\
\midrule
Booking Experience   & 1{,}372 & 3.66 \\
Breakfast            &    788  & 1.44 \\
Breakfast Quality    &    637  & 4.15 \\
Business Center      &    142  & 3.23 \\
Casino               &    208  & 1.36 \\
Loyalty Program      &    607  & 1.57 \\
Nightly Rate         & 2{,}481 & 4.60 \\
Pool                 & 1{,}318 & 2.10 \\
Room Quality         & 2{,}557 & 1.78 \\
Shuttle Service      &    280  & 4.90 \\
\midrule
\textbf{Mean}        & ---     & \textbf{2.88} \\
\bottomrule
\end{tabular}%
}
\end{table}

\section{Robustness \& Sensitivity}
\label{app:sec:robustness}

\subsection{Noise Tolerance}
\label{app:robustness}

Two studies stress-test the system under realistic noise. First, Dirichlet perturbation of $P_{\text{pop}}$ simulates inaccurate population estimates. Second, mislabel sensitivity corrupts both document SI labels and $P_{\text{pop}}$ at once. These jointly reveal how much noise each algorithm can absorb, and where deployment breaks down.

\subsubsection{Dirichlet Perturbation of $P_{\text{pop}}$}

We perturb $P_{\text{pop}}$ with Dirichlet noise at $\varepsilon \in \{0, 0.1, 0.2, 0.5\}$ ($N{=}30$ samples per entity) to simulate estimation error in the population distribution (Figure~\ref{fig:robustness-dirichlet}):

\begin{figure*}[ht!]
\centering
\includegraphics[width=0.92\textwidth]{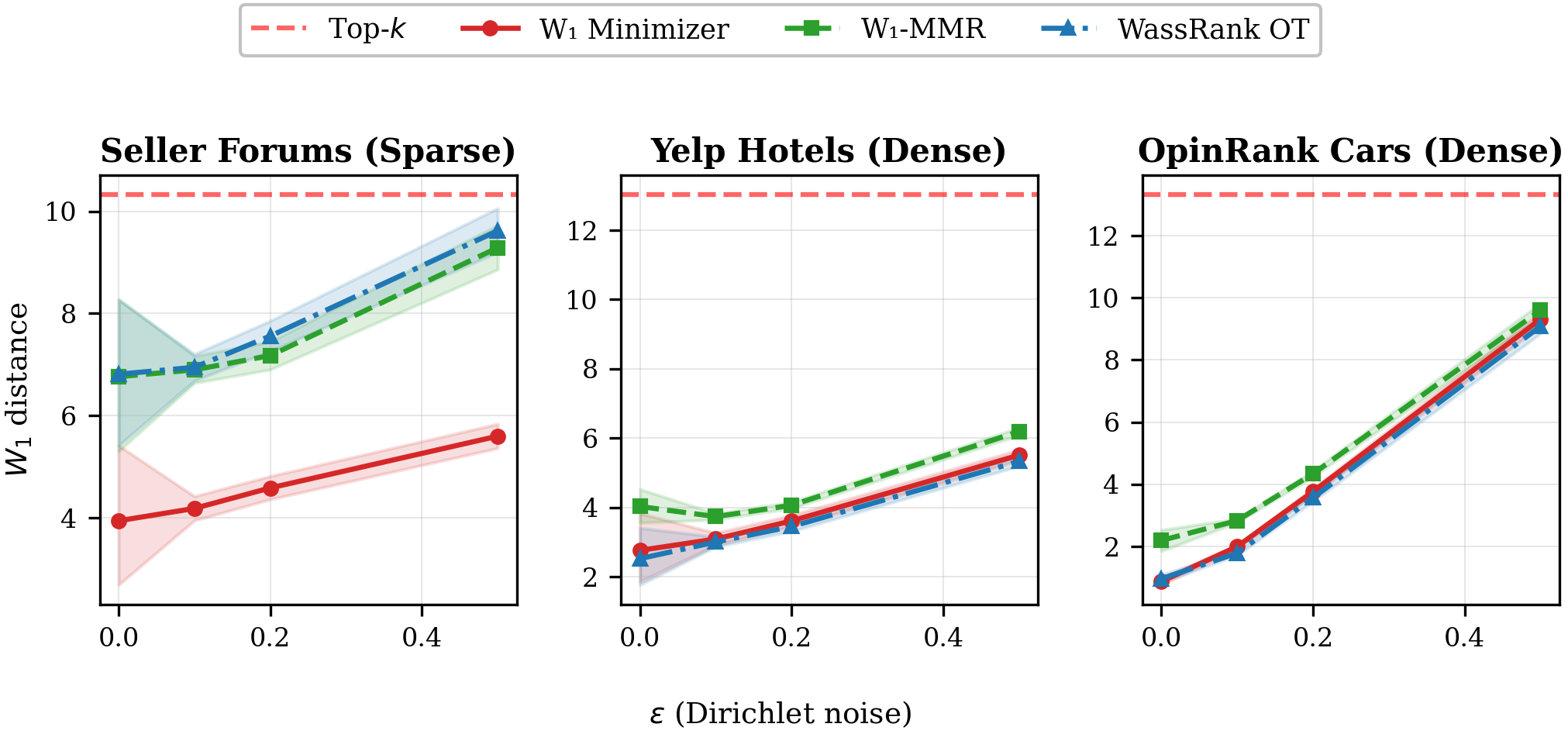}
\caption{Dirichlet perturbation robustness across all three domains. Dashed red line = Top-$k$ baseline. At $\varepsilon{=}0.2$, $W_1$ Minimizer remains 72\% better than Top-$k$ on Yelp.}
\label{fig:robustness-dirichlet}
\end{figure*}

On Yelp, which has a dense candidate pool, the $W_1$ Minimizer at $\varepsilon{=}0.2$ degrades by 31\%, but it still comes in 72\% ahead of Top-$k$. $W_1$-MMR barely moves under the same perturbation (${<}1\%$ degradation). On Amazon Seller Forums the picture is similar in shape: all methods degrade gracefully, and $W_1$ Minimizer at $\varepsilon{=}0.2$ stays 56\% above Top-$k$. OpinRank is the most sensitive of the three because the unperturbed baseline is already tight ($W_1$ Min sits at 0.87 at $\varepsilon{=}0$), so relative degradation looks larger, but even there $\varepsilon{=}0.2$ holds 45\% better than Top-$k$. Across all three domains, $W_1$-MMR degrades the least, which we attribute to its hybrid objective partially shielding it from target noise.

\subsubsection{Mislabel Sensitivity (SI Label Error)}

A harder question: at what SI label error rate does the $W_1$ advantage drop below significance? We corrupt candidate document labels AND $P_{\text{pop}}$ simultaneously, the realistic scenario where labeling errors propagate into the population estimate. Error rates $\in \{0, 0.05, 0.10, 0.20, 0.30, 0.50\}$, $N{=}20$ repetitions per setting (Figure~\ref{fig:robustness-mislabel}). ``Crossover'' marks the error rate at which a method's $W_1$ exceeds the Top-$k$ baseline:

\begin{figure*}[ht!]
\centering
\includegraphics[width=0.90\textwidth]{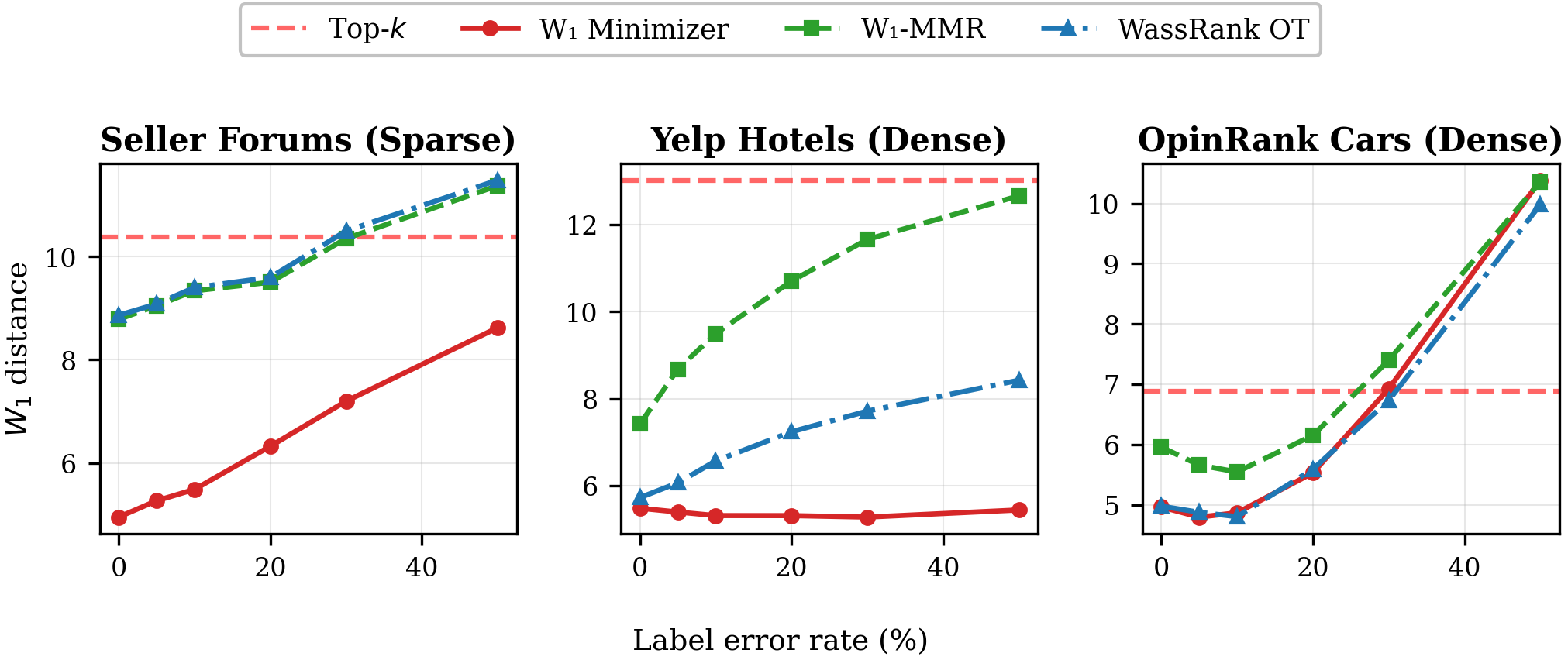}
\caption{Mislabel sensitivity: $W_1$ vs.\ label error rate. Dashed red line = Top-$k$ baseline. $W_1$ Minimizer never crosses Top-$k$ even at 50\% error on Yelp.}
\label{fig:robustness-mislabel}
\end{figure*}

Dense pools are forgiving. $W_1$ Minimizer never crosses the Top-$k$ baseline on Yelp, even at 50\% label error. Sparse pools need more care; use $W_1$-MMR regardless of label quality, as its hybrid objective guarantees a floor.

\subsection{Noise Control}
\label{app:noise-control}

We need to rule out a trivial explanation: maybe any score perturbation induces diversity that looks like opinion-aware signal. So we add 5\% Gaussian noise to retrieval scores and re-rank:

\begin{table}[ht!]
\centering
\small
\caption{Noise control: 5\% Gaussian perturbation of retrieval scores yields negligible $W_1$ improvement (2--5\%, $p > 0.05$), confirming that our 43--89\% distributional improvements ($W_1$-MMR on sparse pools; $W_1$ Minimizer / EG+$W_1$ Min on dense) represent genuine distributional optimization.}
\resizebox{\columnwidth}{!}{%
\begin{tabular}{lccc}
\toprule
\textbf{Method} & \textbf{Seller $W_1$} & \textbf{Yelp $W_1$} & \textbf{OpinRank $W_1$} \\
\midrule
Top-$k$ (baseline) & 10.33 & 13.60 & 13.33 \\
Top-$k$ + 5\% Noise & 10.07 & 13.00 & 12.97 \\
\midrule
\textit{Reduction} & \textit{2.5\%} & \textit{4.4\%} & \textit{2.7\%} \\
\textit{Significance} & \multicolumn{3}{c}{\textit{ns ($p > 0.05$, all domains)}} \\
\midrule
$W_1$ Minimizer & 8.84 & 2.76 & 4.14 \\
\textit{Reduction} & \textit{14.4\%} & \textit{79.7\%} & \textit{69.0\%} \\
\bottomrule
\end{tabular}%
}
\end{table}

Random score perturbation yields 2--5\% $W_1$ change (non-significant), against our 43--89\% distributional improvements on the same pools.

\subsection{Hyperparameter Sensitivity}
\label{app:lambda}
\label{app:n-sensitivity}

\begin{figure*}[ht!]
\centering
\includegraphics[width=0.78\textwidth]{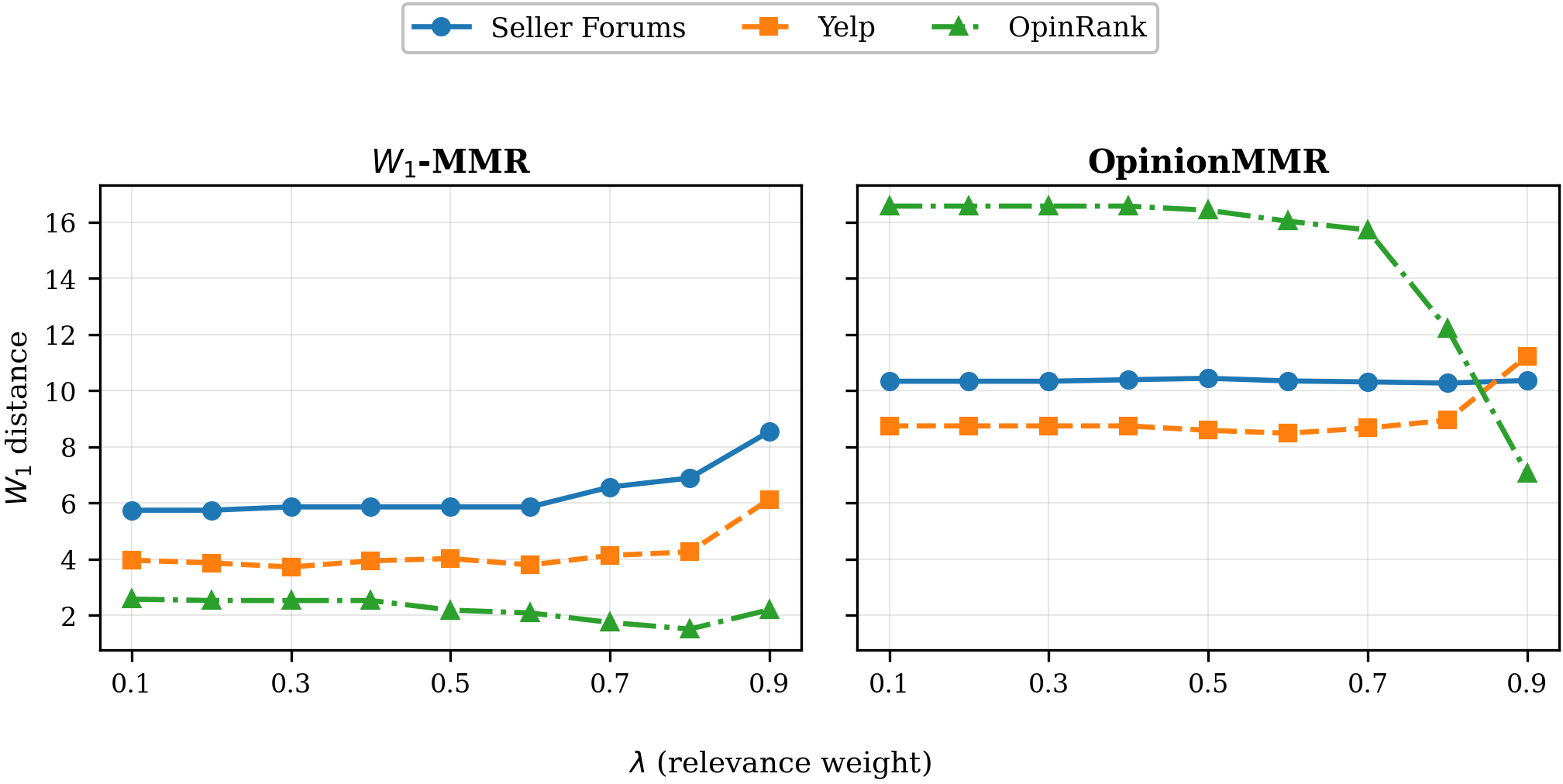}
\caption{$\lambda$-sensitivity for $W_1$-MMR (left) and OpinionMMR (right). $W_1$-MMR is stable at $\lambda \leq 0.5$ and degrades at $\lambda{=}0.9$. OpinionMMR shows higher $W_1$ across all domains with limited sensitivity to $\lambda$.}
\label{fig:lambda-sensitivity}
\end{figure*}

\paragraph{$\lambda$ (relevance--calibration tradeoff).} Hybrid methods ($W_1$-MMR and WassRank OT) each introduce $\lambda$. We sweep $\lambda \in \{0.1, 0.3, 0.5, 0.7, 0.9\}$ across all three domains (Figure~\ref{fig:lambda-sensitivity}). $W_1$-MMR holds steady at $\lambda \leq 0.5$ on Seller Forums and Yelp, then degrades once relevance dominates. OpinRank behaves differently: $\lambda{=}0.7$ is optimal there (denser pools tolerate more calibration weight). WassRank OT is nearly $\lambda$-invariant on dense pools: Yelp gives $W_1 \approx 2.52$ for $\lambda \leq 0.5$ (identical to 3 decimal places) and OpinRank yields $W_1{=}0.95$ with 100\% EM across that range.\footnote{The $\lambda$-sensitivity results use entity-specific FAISS indices (EM=100\%), whereas Table~\ref{tab:main} uses a single shared FAISS index across all OpinRank entities (EM=82.9\%). This accounts for the OpinRank WassRank OT $W_1$ difference (3.87 in Table~\ref{tab:main} vs.\ 0.95 here).} No per-domain tuning needed.

\begin{figure*}[ht!]
\centering
\includegraphics[width=0.78\textwidth]{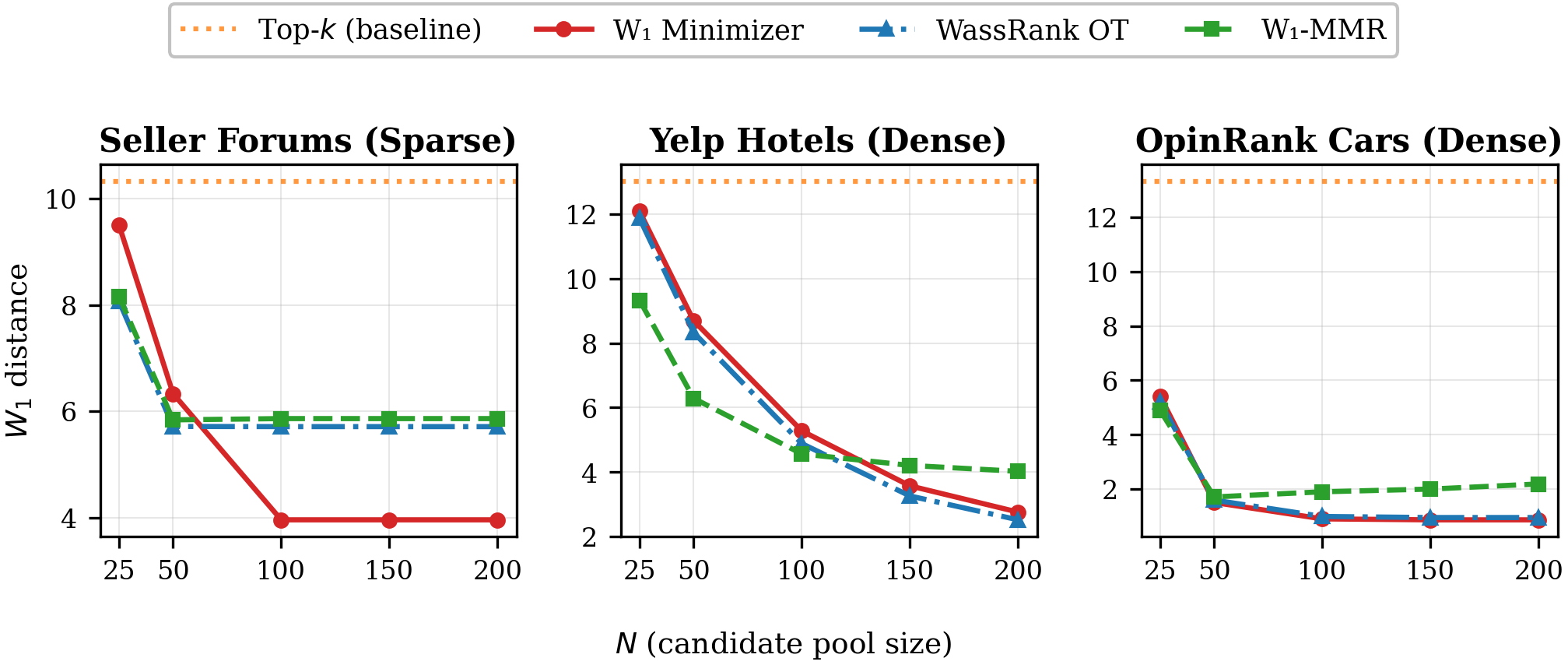}
\caption{$W_1$ vs.\ pool size $N$ ($k{=}20$). $N{=}100$ suffices for dense pools; $W_1$-MMR benefits from larger pools on sparse data.}
\label{fig:n-sensitivity}
\end{figure*}

\paragraph{Pool size $N$.} We vary $N \in \{25, 50, 100, 200\}$ with fixed $k{=}20$ (Figure~\ref{fig:n-sensitivity}). $N{=}100$ suffices for near-optimal performance on dense pools. $W_1$-MMR benefits more from larger pools because it draws from the full (unfiltered) candidate set. On OpinRank, even $N{=}50$ approaches the $N{=}200$ result because entity-specific FAISS indices guarantee most candidates are already entity-matched. On sparse Seller Forums, larger pools become essential for $W_1$-MMR to locate cross-entity documents that fill distributional gaps.

\subsection{Population Estimation and Cold-Start Robustness}
\label{app:ppop-estimation}

We evaluate whether $P_{\text{pop}}$ can be reliably estimated from indexed data and whether the system degrades safely at cold-start. For each entity $e$, the full-corpus distribution $P^*_{\text{pop},e}$ serves as our oracle evaluation target. We sample subsets of $n \in \{10, 25, 50, 100, 250, 500\}$ entity-matched documents, compute $\hat{P}_{\text{pop},e}^{(n)}$, and measure $W_1(\hat{P}_{\text{pop},e}^{(n)}, P^*_{\text{pop},e})$. Crucially, the re-ranker uses $\hat{P}_{\text{pop}}$ but selected evidence is evaluated against $P^*_{\text{pop}}$; this avoids circularity.

\begin{figure}[t]
\centering
\includegraphics[width=\columnwidth]{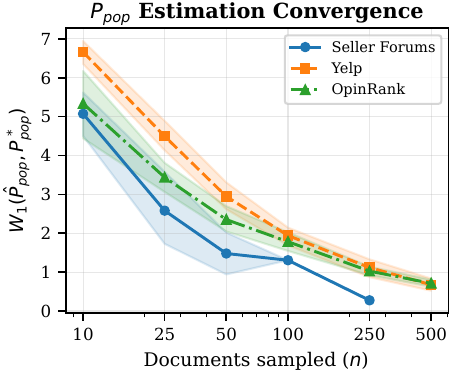}
\caption{$W_1(\hat{P}_{\text{pop}}, P^*_{\text{pop}})$ vs.\ entity-matched documents sampled. With 50--100 documents per entity, estimation error is comparable to or below the re-ranking improvement magnitude itself.}
\label{fig:ppop-convergence}
\end{figure}

\paragraph{Estimation convergence.}
Figure~\ref{fig:ppop-convergence} shows estimation error vs.\ sample size across all three domains. Convergence is rapid. At $n{=}50$, mean $W_1$ between estimated and oracle distributions drops below 3.0; at $n{=}100$, below 2.0. Bootstrap CIs (95\%) narrow monotonically, which indicates stable estimation with modest data.

\paragraph{Cold-start behavior.}
When an entity has few labeled documents, the entity-level $P_{\text{pop}}$ estimate gets noisy. We simulate this by capping observation count at $n \in \{10, 25, 50, 100\}$. Even at $n{=}10$, all estimation strategies beat uncalibrated Top-$k$. Smoothed interpolation with a domain prior ($\tilde{P} = \alpha \hat{P}_e + (1{-}\alpha) P_{\text{domain}}$, $\alpha = n/(n{+}50)$) works best at every count, achieving 61\% $W_1$ reduction at $n{=}10$. Our recommended production rule: use entity-level $P_{\text{pop}}$ when $n \geq 25$ and at least 2 sentiment bins have mass $>0.05$; otherwise fall back to the smoothed domain prior.

\section{Generation Evaluation}
\label{app:sec:generation}

\subsection{Judge Validation \& Positional Bias}
\label{app:judge-validation}

Generation evaluation should not hinge on a single model's quirks; leniency bias and prompt sensitivity are well-documented failure modes~\citep{thakur2025judging}. We validate with five independent model families from different providers: Claude Sonnet 4 (Anthropic), Llama 3.3 70B (Meta), Mistral Large 3 (Mistral AI), Amazon Nova Pro (Amazon), and DeepSeek v3.2 (DeepSeek). Appendix~\ref{app:prompts:judge} contains the full judge prompt with its three evaluation dimensions. Below we confirm positional bias, report directional inter-judge agreement, and present majority-vote consensus.

\subsubsection{Positional Bias}

LLM judges exhibit positional bias, a preference for whichever answer appears first~\citep{wang2024fair,shi2025judging}. We confirm this across all five judges in our panel. Our position-swap protocol converts inconsistent position-dependent preferences into conservative ties, motivating the directional $\kappa_d$ analysis in \S\ref{app:kappa-d}. Bias rates differ across models, producing asymmetric tie distributions:

\begin{table}[ht!]
\centering
\small
\caption{Consensus tie rates across the 5-judge panel. Higher tie rates indicate stronger positional bias (position-swap converts inconsistent preferences to ties).}
\resizebox{\columnwidth}{!}{%
\begin{tabular}{lcc}
\toprule
\textbf{Judge} & \textbf{Consensus Tie Rate ($k{=}5$)} & \textbf{Consensus Tie Rate ($k{=}10$)} \\
\midrule
Claude Sonnet 4 & 46\% & 46\% \\
Mistral Large 3 & 48\% & 61\% \\
DeepSeek v3.2 & 63\% & 71\% \\
Llama 3.3 70B & 73\% & 86\% \\
Amazon Nova Pro & 79\% & 86\% \\
\bottomrule
\end{tabular}%
}
\end{table}

When judges do commit to a preference, they agree directionally 85--100\% of the time.

\subsubsection{Directional Inter-Judge Agreement ($\kappa_d$)}
\label{app:kappa-d}

We want to separate genuine directional disagreement from artifacts of unequal tie rates. So we compute $\kappa_d$ over only those pairs where both judges commit to a non-tie preference:

\begin{table}[ht!]
\centering
\small
\caption{Mean directional $\kappa_d$ per judge vs.\ all others (fairness dimension). Non-Anthropic judges agree at $\kappa_d{\geq}0.95$ at $k{=}10$; at $k{=}5$, $\kappa_d{=}0.88$--$0.93$.}
\resizebox{\columnwidth}{!}{%
\begin{tabular}{lcccc}
\toprule
\textbf{Judge (vs.\ others)} & \textbf{$k{=}5$ $\kappa_d$} & \textbf{$k{=}5$ Agree\%} & \textbf{$k{=}10$ $\kappa_d$} & \textbf{$k{=}10$ Agree\%} \\
\midrule
Claude vs.\ others & 0.68 & 88.2\% & 0.86 & 93.5\% \\
Llama 3.3 vs.\ others & 0.89 & 96.0\% & 0.97 & 98.8\% \\
Mistral L3 vs.\ others & 0.89 & 95.8\% & 0.95 & 97.8\% \\
DeepSeek vs.\ others & 0.93 & 97.4\% & 0.95 & 97.9\% \\
Nova Pro vs.\ others & 0.88 & 95.3\% & 0.94 & 97.3\% \\
\bottomrule
\end{tabular}%
}
\end{table}

All five judges agree directionally at $\kappa_d{=}0.61$--$1.00$. Among the four non-Anthropic judges, pairwise $\kappa_d$ ranges 0.95--1.00 at both $k$ values. Claude's lower $\kappa_d$ (0.68/$k{=}5$, 0.86/$k{=}10$) traces to its lower tie rate: it commits more often and occasionally flags nuances others collapse into ties.

\subsubsection{Majority Vote Results (3/5 Judges Must Agree)}
\label{app:generation}
\label{app:gen-k20}

Under majority vote (3/5 agreement required), calibration methods reach strong significance at $k{=}5$ while Random correctly fails. This is a critical sanity check. At $k{=}10$, fewer approaches survive; context-window averaging attenuates distributional differences. We extend to $k{=}20$ under the same protocol; the scaling regime distinguishes semantic-diversity from distributional methods.

\begin{table*}[ht!]
\centering
\small
\caption{5-judge majority-vote generation results (cross-domain, 156 questions) at $k{\in}\{5,10,20\}$. All calibration methods achieve $p<0.001$ at $k{=}5$; retain significance at $k{=}10$ ($p \leq 0.021$) and $k{=}20$ ($p \leq 0.004$). Random fails at all $k$ values. At $k{=}20$, MMR/OpinionMMR Fair\% collapse (29\%/33\%) as semantic diversity saturates; distributional methods maintain or grow their advantage; WassRank OT rises 70\%$\to$73\%$\to$92\%. Fair\% = W/(W+L). Win\% = W/$N$.}
\label{tab:gen-cross-k10}
\label{tab:gen-cross-k20}
\resizebox{\textwidth}{!}{%
\begin{tabular}{l ccc ccc ccc}
\toprule
& \multicolumn{3}{c}{$k{=}5$} & \multicolumn{3}{c}{$k{=}10$} & \multicolumn{3}{c}{$k{=}20$} \\
\cmidrule(lr){2-4} \cmidrule(lr){5-7} \cmidrule(lr){8-10}
\textbf{Approach} & \textbf{W/L/T} & \textbf{Fair\%} & \textbf{Win\%} & \textbf{W/L/T} & \textbf{Fair\%} & \textbf{Win\%} & \textbf{W/L/T} & \textbf{Fair\%} & \textbf{Win\%} \\
\midrule
Random                 & 16/36/104 & 31\% & 10\% & 12/40/104 & 23\% & 8\%  &  3/26/127 & 10\% & 2\% \\
MMR$^{**}$             & 39/8/109  & 83\% & 25\% & 21/9/126  & 70\% & 13\% &  5/12/139 & 29\% & 3\% \\
OpinionMMR$^{**}$      & 50/11/95  & 82\% & 32\% & 25/8/123  & 76\% & 16\% &  7/14/135 & 33\% & 4\% \\
DPP$^{**}$             & 54/7/95   & 89\% & 35\% & 42/6/108  & 88\% & 27\% & 25/3/128  & 89\% & 16\% \\
\midrule
WassRank OT$^{**}$     & 66/28/62  & 70\% & 42\% & 75/28/53  & 73\% & 48\% & 33/3/120  & \textbf{92\%} & 21\% \\
W$_1$-MMR$^{**}$       & 44/10/102 & 81\% & 28\% & 24/7/125  & 77\% & 15\% & 13/2/141  & 87\% & 8\% \\
W$_1$ Minimizer$^{**}$ & 51/8/97   & 86\% & 33\% & 32/8/116  & 80\% & 21\% & 27/5/124  & 84\% & 17\% \\
\bottomrule
\end{tabular}%
}
\end{table*}

At $k{=}5$ with only 2--3 active sentiment bins, diversity and calibration objectives partly overlap and MMR/OpinionMMR win 83\%/82\% (Table~\ref{tab:gen-cross-k10}). By $k{=}20$ the pool has room for finer-grained proportions and pure semantic-diversity has run out of signal; MMR and OpinionMMR Fair\% drop to 29\% and 33\%. Distributional methods maintain their advantage: WassRank OT rises from 70\% ($k{=}5$) to 73\% ($k{=}10$) to 92\% ($k{=}20$) as budget lets proportional targeting express itself in the answer. DPP is a partial exception: it also grows with $k$ (89\% at $k{=}20$) because its determinantal spread lands one representative per bin when there is enough budget.

Domain decomposition at $k{=}10$ reveals a clear density dependency. On dense Yelp (60 questions, 10 aspects), all calibration methods achieve 86--97\% fairness decided rate; $W_1$ Minimizer hits 97\% (28W/1L). Dense OpinRank (60 questions, 10 entities) confirms propagation: DPP leads at 100\% (19W/0L), $W_1$ Minimizer at 94\% (16W/1L), and 5 of 8 approaches reach $p<0.001$. Sparse Seller Forums (36 questions, 6 entities, EM${\sim}$40\%) is the outlier: only OpinionMMR achieves significance (10W/2L, 83\%, $p{=}0.019$); $W_1$ Minimizer and $W_1$-MMR hover near coin-flip (54--55\%), confirming that sparse pools cap document quality regardless of re-ranking strategy. Random fails on all three domains (13--53\%), validating the sanity check.

\subsubsection{Judge Prompt Sensitivity}
\label{app:prompt-ablation}

Our main fairness criterion conflates two signals: \emph{proportional accuracy} and \emph{viewpoint coverage}. We disentangle them by re-judging all answer pairs under two contrasting definitions (Prompt~A and Prompt~B in Appendix~\ref{app:prompts:judge}), same 5-judge panel with position-swap~\citep{zheng2024judging} and 3/5 majority vote:

\begin{itemize}
\item \textbf{Prompt~A (Proportional Accuracy):} Judges receive the ground-truth $P_{\text{pop}}$ and are asked which answer's emphasis better matches the actual distribution. An answer over-representing a 10\% minority relative to a 70\% majority is penalized as misleading.
\item \textbf{Prompt~B (Viewpoint Coverage):} Judges are asked which answer covers more distinct viewpoints, including rare ones. Giving disproportionate space to minority views is rewarded.
\end{itemize}

INFORMATIVENESS and GROUNDEDNESS dimensions remain identical across both prompts, serving as controls.

\begin{table}[ht!]
\centering
\small
\caption{Judge prompt ablation: fairness win rate (\%) under proportional vs.\ coverage definitions (3 domains, 156 questions per approach). $W_1$ Minimizer dominates under proportional accuracy; the gap narrows under coverage. Stratified sampling (oracle) confirms the Prompt~A criterion tracks the ground-truth distribution, not any specific retrieval strategy --- $W_1$ Minimizer matches oracle Fair\% within 1\,pp at $k{=}5$.}
\label{tab:prompt-ablation}
\resizebox{\columnwidth}{!}{%
\begin{tabular}{l cc cc}
\toprule
& \multicolumn{2}{c}{$k{=}5$} & \multicolumn{2}{c}{$k{=}10$} \\
\cmidrule(lr){2-3} \cmidrule(lr){4-5}
\textbf{Approach} & \textbf{Prop.\ (A)} & \textbf{Cov.\ (B)} & \textbf{Prop.\ (A)} & \textbf{Cov.\ (B)} \\
\midrule
Stratified (oracle) & 35\% & --- & --- & --- \\
$W_1$ Minimizer & 34\% & 38\% & 30\% & 27\% \\
DPP & 26\% & 35\% & 24\% & 25\% \\
$W_1$-MMR & 23\% & 36\% & 18\% & 17\% \\
OpinionMMR & 21\% & 30\% & 13\% & 19\% \\
MMR & 13\% & 26\% & 10\% & 13\% \\
Random & 16\% & 13\% & 12\% & 7\% \\
\bottomrule
\end{tabular}%
}
\end{table}

\begin{table}[ht!]
\centering
\small
\caption{Control dimensions remain stable across prompt variants ($k{=}5$, $\Delta$ = Prompt~B $-$ Prompt~A). Shifts $\leq$4\,pp confirm the fairness prompt change does not contaminate other evaluation axes.}
\label{tab:prompt-ablation-controls}
\resizebox{\columnwidth}{!}{%
\begin{tabular}{l cccc}
\toprule
\textbf{Approach} & \textbf{Info\% (A)} & \textbf{Info\% (B)} & \textbf{Grnd\% (A)} & \textbf{Grnd\% (B)} \\
\midrule
$W_1$ Minimizer & 44\% & 40\% & 28\% & 24\% \\
DPP & 38\% & 40\% & 27\% & 24\% \\
Random & 19\% & 18\% & 15\% & 8\% \\
\bottomrule
\end{tabular}%
}
\end{table}

Three patterns emerge. (1)~Under proportional accuracy, $W_1$ Minimizer leads DPP by 8/6\,pp at $k{=}5$/$10$ and sits within 1\,pp of the stratified oracle (35\%), confirming Prompt~A tracks ground-truth $P_{\text{pop}}$ rather than any retrieval strategy. (2)~Under viewpoint coverage, the gap narrows to 3/2\,pp; DPP's diversity edge is real but modest once $W_1$'s calibration already covers most bins. (3)~Inter-judge agreement rises under Prompt~B (Llama--Mistral $\kappa$: 0.39$\to$0.62 at $k{=}5$), suggesting viewpoint counting is more objectively evaluable without ground truth. Our original conflated criterion thus understates $W_1$'s advantage at its design objective.

\subsection{Single-Judge Metric Isolation (Generation)}
\label{app:metric-isolation}

The 5-judge majority-vote protocol is conservative --- positional-bias-driven ties inflate the tie column. To isolate the ordinal metric's end-to-end contribution with more statistical power, we ran a controlled single-judge (Claude Sonnet 4) position-controlled pairwise evaluation between $W_1$ Minimizer and KL(JS) Minimizer at $k{=}5$ (Table~\ref{tab:metric-gen}). Same algorithm, same entity-filtered pool; only the distance function differs.

\begin{table}[ht!]
\centering
\small
\caption{$W_1$ Minimizer vs.\ KL(JS) Minimizer, generation evaluation at $k{=}5$ (156 queries, single-judge Claude Sonnet 4, position-controlled, blind pairwise vs.\ Top-$k$). $W_1$ leads KL(JS) by +7\,pp cross-domain; both significantly beat Top-$k$ ($p{<}0.001$).}
\label{tab:metric-gen}
\resizebox{\columnwidth}{!}{%
\begin{tabular}{llrrrr}
\toprule
\textbf{Domain} & \textbf{Method} & $N$ & \textbf{W/L/T} & \textbf{Fair\%} & \textbf{Win\%} \\
\midrule
Seller Forums & $W_1$ Minimizer   & 36  & 10/9/17  & \textbf{53\%} & \textbf{28\%} \\
              & KL (JS) Minimizer & 36  &  7/10/19 & 41\%          & 19\% \\
\midrule
Yelp          & $W_1$ Minimizer   & 60  & 39/4/17  & \textbf{91\%} & \textbf{65\%} \\
              & KL (JS) Minimizer & 60  & 31/8/21  & 79\%          & 52\% \\
\midrule
OpinRank      & $W_1$ Minimizer   & 60  & 29/8/23  & 78\%          & 48\% \\
              & KL (JS) Minimizer & 60  & 35/10/15 & 78\%          & \textbf{58\%} \\
\midrule
Cross-domain  & $W_1$ Minimizer   & 156 & 78/21/57 & \textbf{79\%} & \textbf{50\%} \\
              & KL (JS) Minimizer & 156 & 73/28/55 & 72\%          & 47\% \\
\bottomrule
\end{tabular}%
}
\end{table}

$W_1$ leads KL(JS) by +7\,pp Fair\% cross-domain; the retrieval-side advantage (Appendix~\ref{app:kl-ablation}) does not invert at generation. Yelp shows the largest gap (91\% vs.\ 79\%); Seller Forums the same direction at smaller magnitude (53\% vs.\ 41\%). OpinRank is a partial exception (Fair\% ties; KL(JS) edges Win Rate 58\% vs.\ 48\%) --- its near-uniform $P_{\text{pop}}$ compresses ordinal distances, leaving the ordinal ground cost less to do.

\end{document}